\documentclass[reprint,aps,pra,twocolumn,superscriptaddress,twocolumn,showpacs,longbibliography]{revtex4-1}
\usepackage{bm,amsmath,amssymb,amsfonts}
\usepackage{color}
\usepackage{dcolumn}
\usepackage{graphicx}
\usepackage{float}
\usepackage{hyperref}
\hypersetup{colorlinks=true,citecolor=blue,urlcolor=blue,linkcolor=blue}

\usepackage{times}
\usepackage{subfigure}

\begin{document}

\title{The time-domain Landau-Zener-St\"{u}ckelberg-Majorana interference in optical lattice clock}
\author{Wei-Xin Liu}
\affiliation{Institute of Theoretical Physics and Department of Physics, State Key Laboratory of Quantum Optics and Quantum Optics Devices, Collaborative Innovation Center of Extreme Optics, Shanxi University, Taiyuan 030006, China}
\author{Tao Wang}
\thanks{corresponding author: tauwaang@cqu.edu.cn}
\affiliation{Department of Physics, and Center of Quantum Materials and Devices, Chongqing University, Chongqing, 401331, China}
\affiliation{Chongqing Key Laboratory for Strongly Coupled Physics, Chongqing, 401331, China}
\author{Xue-Feng Zhang}
\affiliation{Department of Physics, and Center of Quantum Materials and Devices, Chongqing University, Chongqing, 401331, China}
\affiliation{Chongqing Key Laboratory for Strongly Coupled Physics, Chongqing, 401331, China}
\author{Wei-Dong Li}
\affiliation{Shenzhen Key Laboratory of Ultraintense Laser and Advanced Material Technology, Center for Advanced Material Diagnostic Technology, and College of Engineering Physics, Shenzhen Technology University, Shenzhen, 518118, China}

\begin{abstract}
The interference between a sequence of Landau-Zener (LZ) transitions can produce Rabi oscillations (LZROs). This phenomenon is a kind of time-domain Landau-Zener-St\"{u}ckelberg-Majorana (LZSM) interference. However, experimental demonstrations of this LZSM interference induced Rabi oscillation are extremely hard due to the short coherence time of the driven quantum system. Here, we study theoretically the time-domain LZSM interference between the clock transition in one-dimensional (1D) $^{87}$Sr optical lattice clock (OLC) system. With the help of both the adiabatic-impulse model and Floquet numerical simulation method, the LZROs with special step-like structure are clearly found both in fast- and slow-passage limit in the real experiment parameter regions. In addition, the dephasing effect caused by the system temperature can be suppressed with destructive interference in the slow-passage limit. Finally, we discuss the possible Bloch-Siegert shift while the pulse time is away from the half-integer and integer periods. Our work provides a clear roadmap to observe the LZROs on the OLC platform.
\end{abstract}

\maketitle

\section{INTRODUCTION}
The transition between two energy states at an avoided level crossing is known as the Landau-Zener (LZ) transition. When the system is modulated to traverse the avoided crossing points back and forth, the phase collected between two levels during the transition may result in constructive or destructive interference, which is often referred to Landau-Zener-St\"{u}ckelberg-Majorana (LZSM) interference \cite{Landau,Zener,Stuck,Major}. The LZSM interference is analyzed in a variety of physical systems \cite{JRPrl1995,Science2005,MSPrl2006,LCPra2010,ZQPra2008,CSEPra2009,GZSPrb2011,JHZPra2014,FFPrl2014,HYLPra2019,JLPra2020}, and it provides a possible approach to control the quantum states and build quantum gates for quantum computing \cite{YTPrl1998,KSPrb2004,LGNP2012,GCNc2013,FGPra2003}. In particular, the LZSM interference process can be used for designing an interferometer if it can be coupled to the environment \cite{ShytovEPJB2003,HuangPRX2011}.

On the other hand, the interference between multiple successive LZ transitions can lead to periodic oscillations of the population in different energy levels \cite{Physrep2010,PraAshhab2007}, which is similar to Rabi oscillations \cite{Rabi1937}. Although this time-domain oscillations of LZSM interferences are usually named as LZ-Rabi oscillations (LZROs), it presents more rich and distinctive phenomena \cite{DuPrl2014,NeilingerPRB2016} in comparison with the conventional Rabi oscillations. 
Meanwhile, they can also be used to analyze the decay rate of the quantum system \cite{DuPrl2014,HDRPrb2012}. In comparison with LZSM interference which have just been observed in various experiments \cite{Science2005,MSPrl2006,LCPra2010,FFPrl2014,NcLJ2013}, the time-domain LZROs are hardly realized because the coherence time of most related systems is practically shorter than requirement for a long sequence of LZ processes. It has only been demonstrated by making use of the nitrogen-vacancy (NV) center in diamond \cite{DuPrl2014}, however, never observed in the atomic system.

The OLC system, based on the doubly forbidden clock transition between $^1$S$_0$($|g\rangle$) and $^3$P$_0$($|e\rangle$), is one of the most appropriate candidates for observing LZROs. The long lifetime of the excited clock state helps to keep the coherence and can also reduce the atom loss caused by the spontaneous emission \cite{LudRmp2015,BoydScience2006}. On the other hand, based on the state-insensitive optical trap with ``magic" wavelength, the fairly clean OLC system can finely tailor the target Hamiltonian \cite{MLWPrl2016,SKNature2017}. Then, the only task left would be periodically modulating the transition frequency.

Recently, it has been realized in one dimensional $^{87}$Sr OLC system by modulating the lattice frequency around the ``magic" wavelength \cite{YinCPL2021,wangprl}. Due to the Doppler effect, the dynamics of internal clock transition is governed by a photon-assisted LZSM Hamiltonian \cite{Physrep2010,NcLJ2013,Report2017}
\begin{equation}
\hat{H}_{\rm LZSM}^{n_z,n_x}(t) = \frac{\hbar}{2}[\delta + A\omega_s\cos(\omega_st)]\hat{\sigma}_z + \frac{\hbar}{2}g_{n_z,n_x}\hat{\sigma}_x, \label{0}
\end{equation}
where $\delta$ is the frequency difference between the two-level energy gap and clock laser, $\omega_s$ is the driving frequency, $A$ is the renormalized dimensionless driving amplitude, and $g_{n_z,n_x}$ is the clock laser coupling strength of the external energy level $|n_z,n_x\rangle$ \cite{YinCPL2021,Pra2009Ye}. As demonstrated in previous work \cite{YinCPL2021}, more than ten Floquet sidebands are clearly observed with stable spectroscopy sensitivity, and it can be well understood under the resolved side-band approximation (RSBA) in the parameter region $\omega_s \gg g_{n_z,n_x}$. Meanwhile, the dynamics of the Floquet sidebands show smooth sinusoidal behavior which follows the Rabi oscillations with effective coupling strengths modified by Bessel functions. However, the characteristics of LZROs are not found, such as the population oscillations with step-like structure.

In this manuscript, we study theoretically the time-domain LZSM interference between the clock states of the OLC system. Our proposal is based on the experimental realized driven OLC system with Hamiltonian Eq.(\ref{0}) \cite{YinCPL2021}. 
Different from previous paper \cite{YinCPL2021} working on the RSBA, we enter into the parameter regions beyond RSBA and make use of the Floquet theory to solve the system and get accurate numerical results. Meanwhile, we also give an analytic expression and qualitative explanation with the analytic approach -- adiabatic-impulse model (AIM) \cite{BDPra2006,Physrep2010} to describe the time evolution of the system. The LZSM interference between the clock states is studied in both fast and slow passage limits, and the temperature effect is well treated. The AIM results agree well with the numerical results from the Floquet approach, and the time-domain LZROs are found in both limit.In the slow-passage limit, we found the analytic expression of coarse-grained oscillation with effective Rabi frequency is also suitable for the half-integer period in the adiabatic basis. At last, the fixing of zero detuning in the real experiment parameter region is analyzed, so that
the time-domain LZSM interference can be possibly experimental observed in the driven OLC platform under proper chosen parameter region \cite{YinCPL2021}.

The manuscript is organized as follows. In Sec.II, we briefly review the Landau-Zener problem and related adiabatic-impulse model method. In Sec.IIIA, we discuss the Landau Zener transition in the first half period. In Sec.IIIB, we turn to the LZSM interference at one whole period with two avoided crossing points. In Sec.IIIC, we consider the multi-periods in both slow- and fast-passage limits, and the time-domain LZROs are observed with effective Rabi frequency matches well with the analytic results. In Sec. IV, we discuss the frequency shift during the time-domain LZROs. We make a conclusion in Sec.V. A brief description of the model of the driven OLC system and the Floquet approach are introduced in Appendix A and B, respectively.


\begin{figure}[t]
	\centering
	\includegraphics[width=0.5\linewidth]{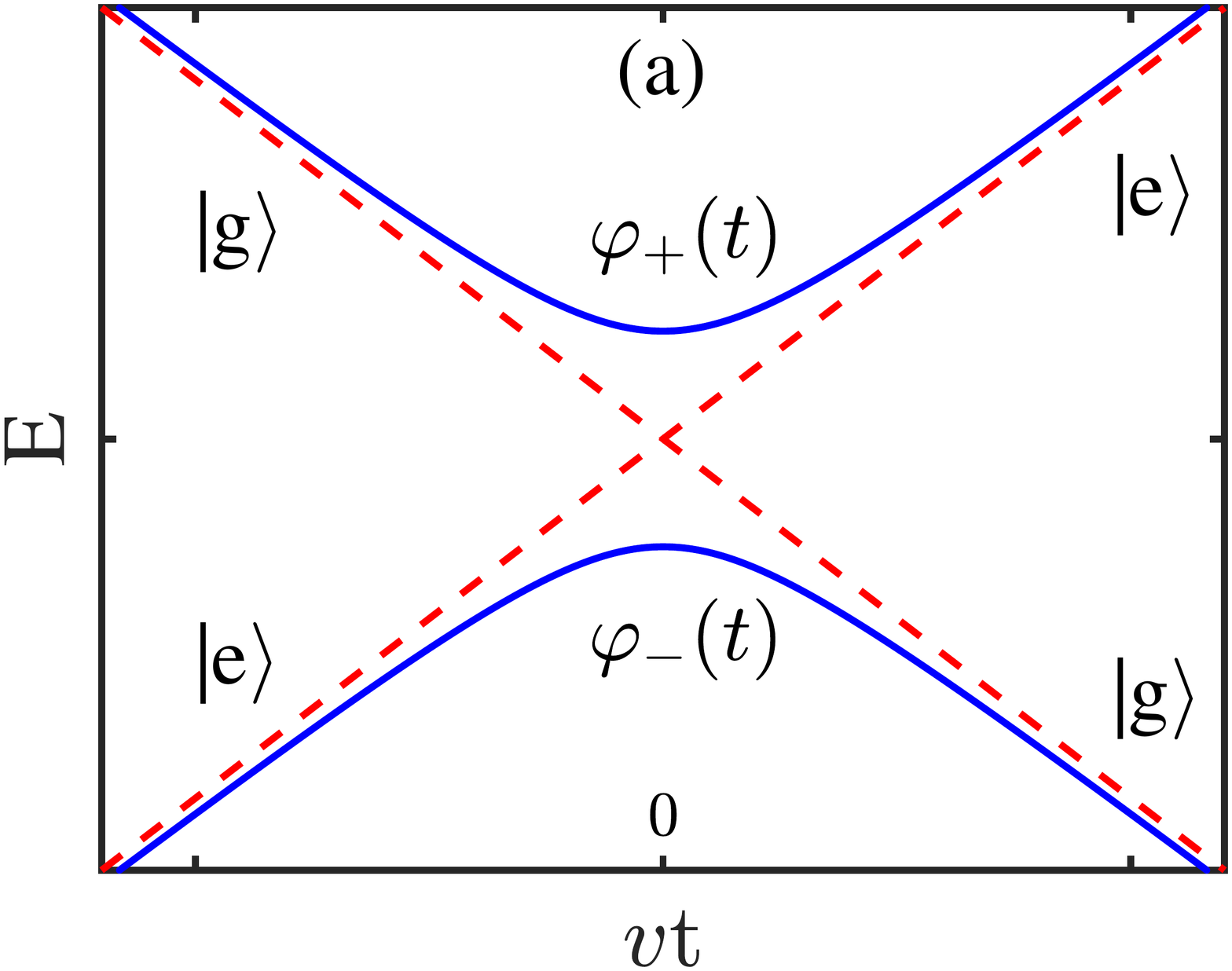}\hfill
	\includegraphics[width=0.5\linewidth]{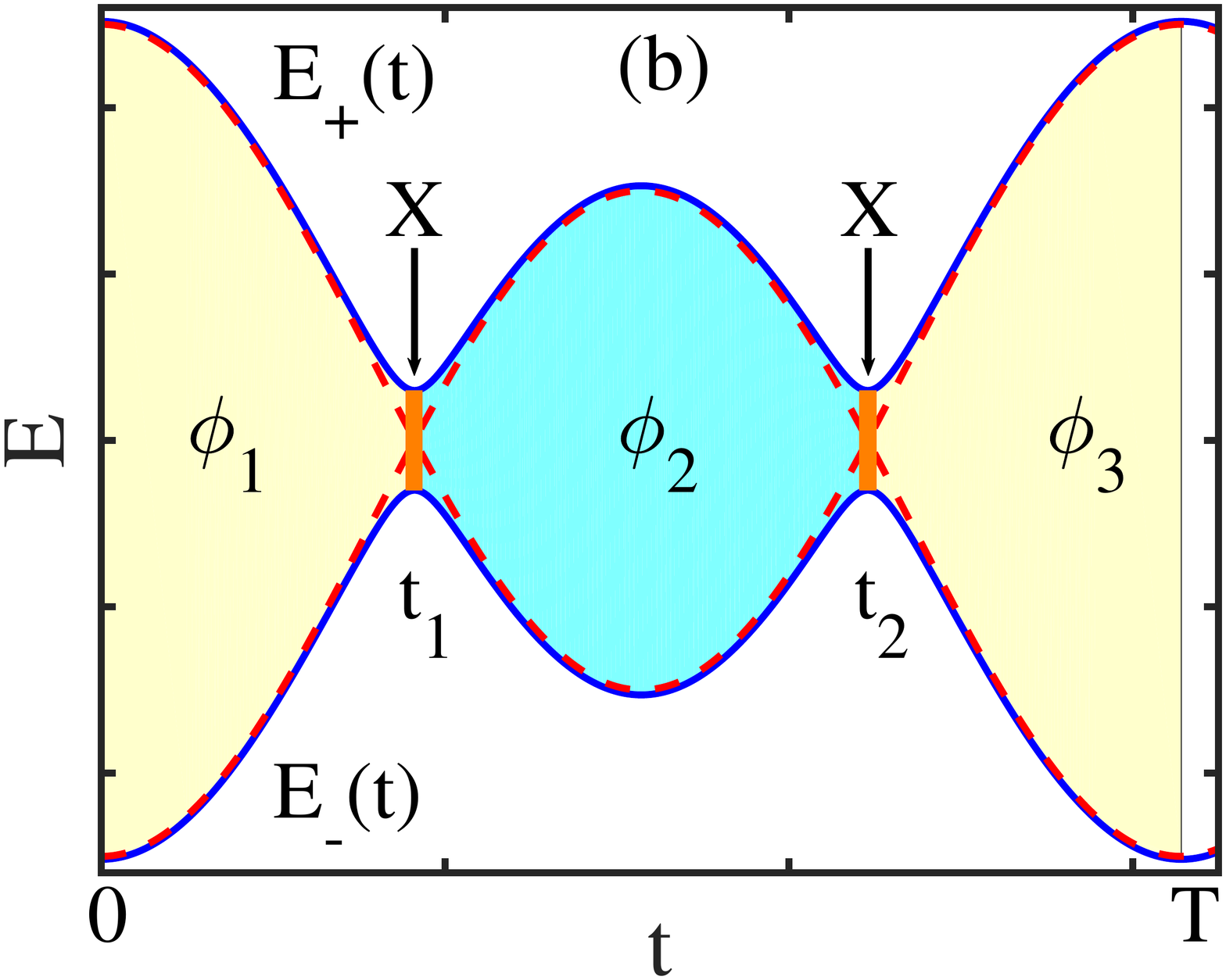}\newline
	\caption{The energy sketch of the driven two-level system. (a) Energy levels versus the driving field $\varepsilon (t)=vt$. The two solid blue curves represent the adiabatic energy levels $\varphi_{\pm}$, and the red dashed lines show the crossing diabatic levels $|e\rangle$ and $|g\rangle$ . (b) Time evolution of the energy levels during one period. At the avoided crossing points, the system undergoes LZ transitions described by unitary transfer matrix $X$, which are denoted by the orange lines with arrows. The colored areas indicate the phases $\phi_{1,2,3}$ collected during the adiabatic evolution. The diabatic energy levels $\pm\varepsilon (t)/2$ are shown by the red dashed lines.}%
	\label{Fig2}%
\end{figure}

\section{LANDAU-ZENER PROBLEM}

\subsection{Model}
The Landau-Zener problem can be described with generic LZSM-Hamiltonian
\begin{equation}
\hat{H}_{\rm LZSM}(t)=\frac{\hbar}{2}\varepsilon(t)\hat{\sigma}_z+\frac{\hbar}{2}g\hat{\sigma}_x,     \label{a1}
\end{equation}
where $\varepsilon(t)$ is the time-dependent longitudinal driving, $g$ is a constant coupling, and the eigenstates of Pauli matrix $\hat{\sigma}_z$ are also called \textit{diabatic states} or basis ${|g\rangle, |e\rangle}$. On the other hand, if we solve the instantaneous eigenfunction $\hat{H}_{\rm LZSM}(t)|\varphi(t)\rangle=E(t)|\varphi(t)\rangle$, we can obtain the instantaneous eigenvalues or eigen-energies $E_{\pm}(t)=\pm\frac{\hbar}{2}\sqrt{\varepsilon^2(t)+g^2}$ with instantaneous eigen-states:
\begin{align}
|\varphi_+(t)\rangle & = \sin \frac{\theta(t)}{2} |g\rangle + \cos \frac{\theta(t)}{2} |e\rangle ,  \nonumber\\
|\varphi_-(t)\rangle & = \cos \frac{\theta(t)}{2} |g\rangle - \sin \frac{\theta(t)}{2} |e\rangle ,                  \label{a2}
\end{align}
where $\theta(t) = \arctan(g/\varepsilon(t))$, and $|\varphi_{\pm}(t)\rangle$ are also named as \textit{adiabatic basis}. When $g=0$, the energy gap closes at the zero points of $\varepsilon(t)$. After linearly expanding $\varepsilon(t)$ around the zero point, it can be approximately written as $vt$, where $v$ is the sweep rate. As shown in Fig. {\ref{Fig2}}(a) in negative $t$ region, the adiabatic states $|\varphi_{+}(t)\rangle$ and $|\varphi_{-}(t)\rangle$ are same as diabatic state $|g\rangle$ and$|e\rangle$. In contrast, they will exchange when passing the energy crossing point (zero point) and entering the positive $t$ region.

When $g$ is nonzero, the two adiabatic states display an avoided crossing with the gap $\hbar g$ at $t=0$. When far away from the avoided crossing point (means when $|\varepsilon(t)|\gg g$), the adiabatic states coincide asymptotically with the diabatic states (see Fig. {\ref{Fig2}}(a)). In the LZ model, one can exactly get the final quantum state for arbitrary initial state by expressing the Schr\"{o}dinger equation in terms of the Weber functions \cite{Zener,Physrep2010}. Assuming the system is in the ground state $|g\rangle$ at initial time $t_i=-\infty$, then the probability in $|g\rangle$ at final time $t_f=+\infty$ is exactly given by the famous LZ formula \cite{Landau,Zener,Stuck}
\begin{equation}
P_{\rm LZ}=\exp(-2\pi\chi)                     \label{a3}
\end{equation}
with $\chi=g^2/(4v)$ called the adiabatic parameter. In the fast-passage limit ($\chi\ll 1$), $P_{\rm LZ} \approx 1$  means the sweep rate is so fast that the system evolves along the diabatic state. In the opposite case -- slow-passage limit ($\chi\gg 1$), $P_{\rm LZ} \approx 0^+$ indicates the system follows the adiabatic path.

\subsection{Adiabatic-impulse model}
In the LZ model, the adiabatic states change rather rapidly around the region of the avoided crossing but approximately keep constant when far from that region. Thus, the evolution of the system can be approximately treated as experiencing a non-adiabatic transition just at the avoided crossing point ($t=0$), apart from which the system undergoes a free adiabatic time-evolution, and it is called adiabatic-impulse model (AIM).
The adiabatic evolution can be described with help of the unitary evolution matrix in the adiabatic basis
\begin{equation}
U_\phi = e^{-i\phi(t_f,t_i)\hat{\sigma}_z},                                                            \label{a4}%
\end{equation}
where $\phi(t_f,t_i)$ is the dynamical phase accumulated from initial time $t_i$ to final time $t_f$%
\begin{equation}
\phi(t_f,t_i) = \frac{1}{2\hbar} \int_{t_i}^{t_f} [E_+(t) - E_-(t)]\,dt.                                 \label{a5}%
\end{equation}
Meanwhile, the instantaneous non-adiabatic transition is approximately governed by the unitary time-independent matrix \cite{Physrep2010,MSPrl2006,Report2017}
\begin{equation}
X=
\begin{pmatrix}
\sqrt{1-P_{\rm LZ}}\exp(-i\varphi) & \sqrt{P_{\rm LZ}}\\
-\sqrt{P_{\rm LZ}}               & \sqrt{1-P_{\rm LZ}}\exp(i\varphi)%
\end{pmatrix},                                                                                       \label{a6}%
\end{equation}
where $\varphi = -\pi/4 + \chi(\ln\chi-1) + \arg[\Gamma(1-i\chi)]$, and $\Gamma$ is the Gamma function. $\varphi$ approaches to $-\pi/4$ in fast-passage limit and to $-\pi/2$ in slow-passage limit. Then, the dynamics of the Landau-Zener problem from initial time $t_i<0$ to final time $t_f>0$ can be proximately in a description of the AIM with the combined evolution matrices $U_{\phi(t_f,0)}XU_{\phi(0,t_i)}$.

In the AIM, the duration time of LZ transition, which is called the LZ jump time, at the avoided crossing point is treated as zero. In order to check the goodness of the approximation, the LZ jump time is defined as $t_{\rm LZ}=P(\infty)/P'(0)$, where $P(\infty)$ is the asymptotic value of the LZ transition, and $P'(0)$ is time derivative of the transition probability at the crossing point $t=0$ \cite{VitanovPra1999}. In the diabatic basis, it proves the following relation as $t_{\rm LZ}\sim 2\sqrt{\pi/v}$ for $\chi \ll 1$ and $t_{\rm LZ}\sim 2g/v$ for $\chi \gg 1$ \cite{VitanovPra1999,GTPra2010}. Apparently, the AIM requires LZ jump time to be sufficiently shorter than the time interval of successive two LZ transitions.

\begin{figure}[t]
	\centering
	\includegraphics[width=1\linewidth]{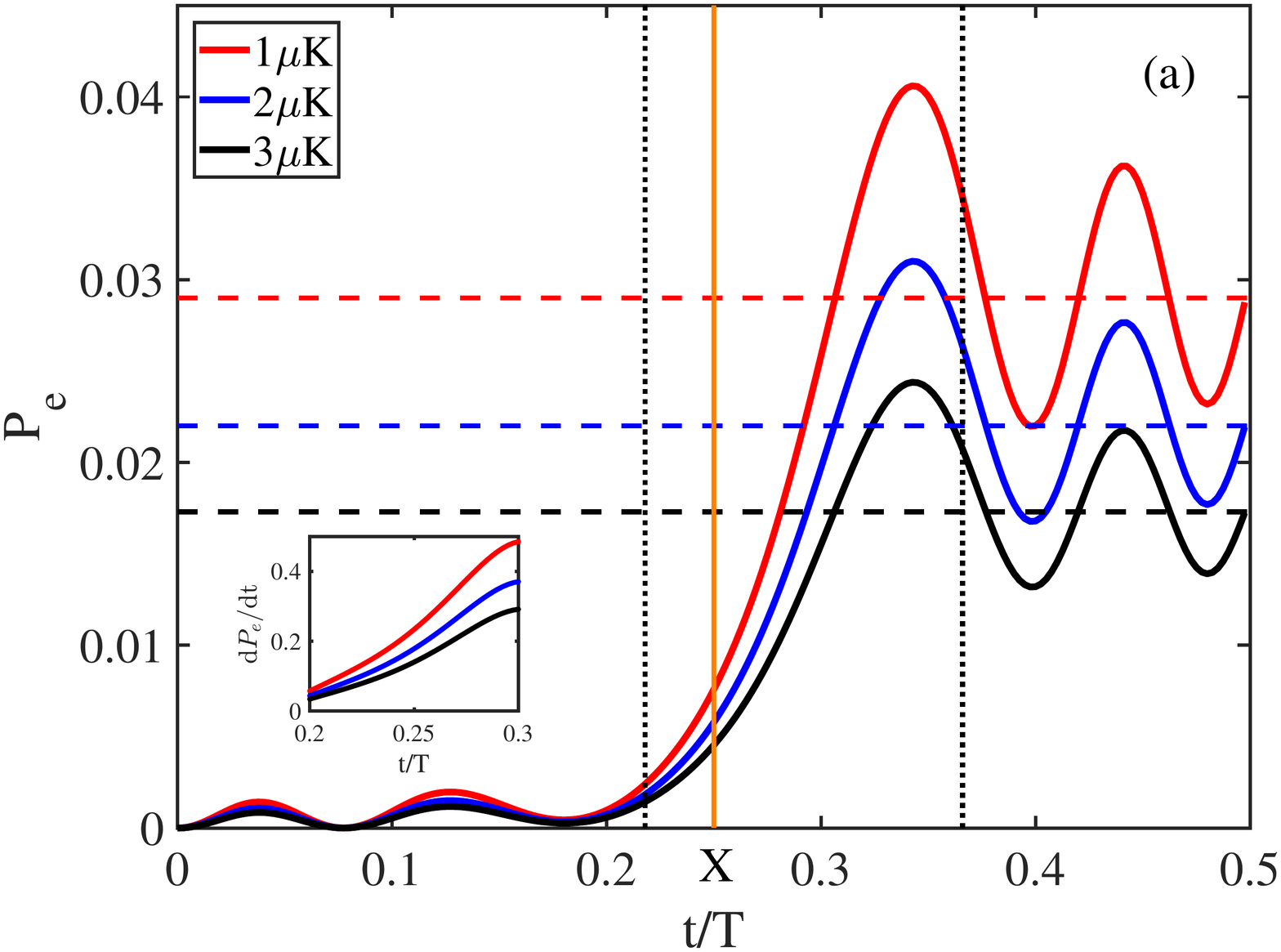}\hfill
	\includegraphics[width=1\linewidth]{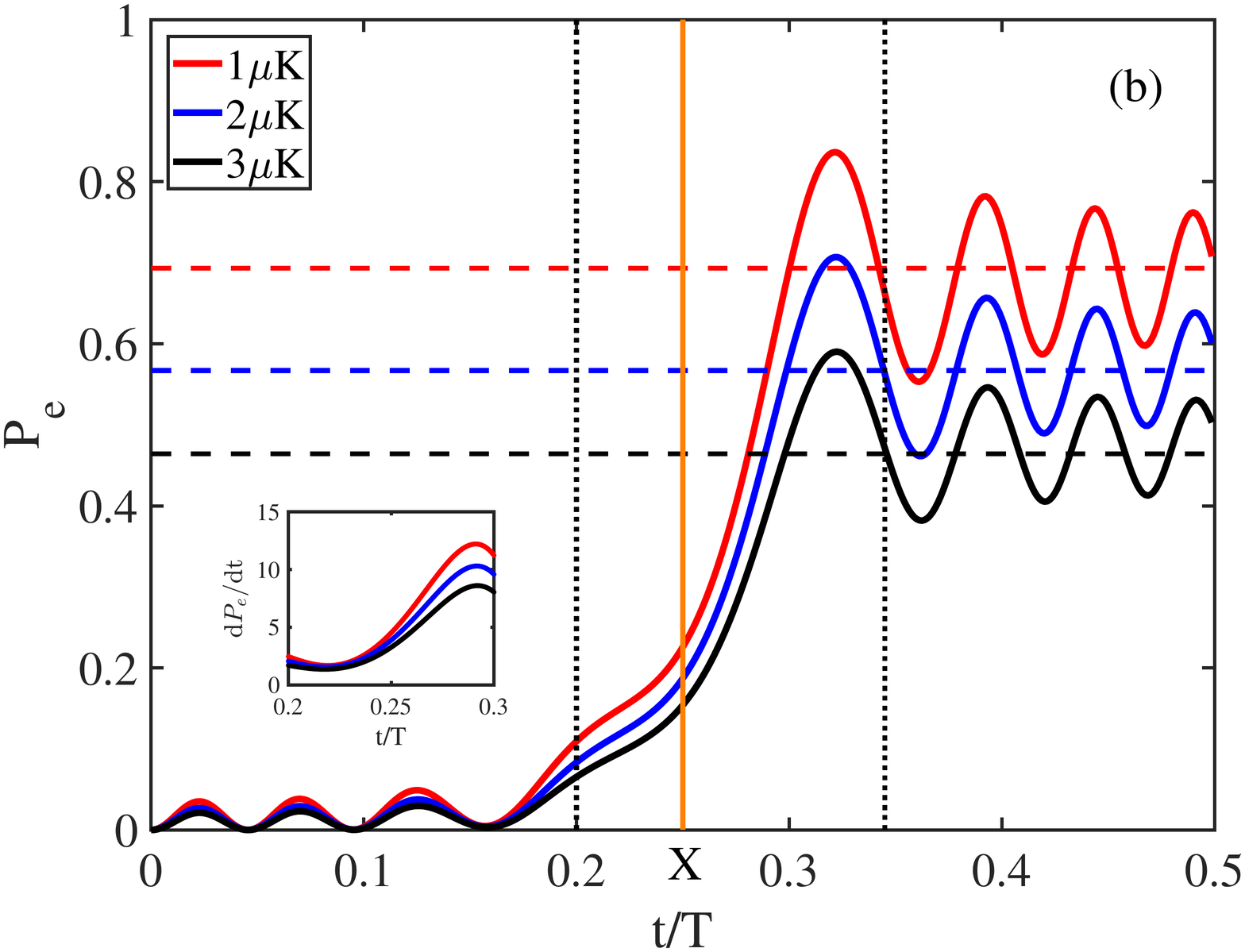}\\
	\caption{LZ transition probability in half driving period. The population probability of state $|e\rangle$ is plotted as a function of time with zero detuning in (a) fast-passage limit with $A=13.3$, $g/\omega_s=0.6$ and (b) slow-passage limit with $A=21.8$, $g/\omega_s=5$. In both figures, the solid lines are numerical results from the Floquet approach, and the dashed lines are the asymptotic LZ transition probability $1-\langle P_{\rm LZ}\rangle_T $. The red, blue, black (solid and dashed) lines corresponding to the results in the temperature 1$\mu$K, 2$\mu$K, 3$\mu$K, respectively. The time regions between the black dotted lines indicate the jump time of LZ transition $\langle t_{\rm LZ} \rangle_T$, which almost remains unchanged in the different temperatures above. The orange lines indicate the position of non-adiabatic X-transition. The insets present the first derivatives of $P_e$ versus time.}%
	\label{Fig3}%
\end{figure}

\section{LZSM interference in OLC}
\subsection{LZ Transition in Half Period}
The OLC system under longitudinal periodical modulation in description of Hamiltonian Eq.(\ref{0}) can be taken as two LZ transitions in one period $T=2\pi/\omega_s$ with $\varepsilon(t)=\delta+A\omega_s\cos(\omega_st)$ and $g=g_{n_z,n_x}$.
When the driving frequency $\omega_s$ is larger than $\delta/A$, the $\varepsilon(t)$ achieves two zero points at $t_1 = \arccos[-\frac{\delta}{A\omega_s}]/\omega_s +n T$ and $t_2 = T-t_1+n T$, where $n\in \mathbb{Z}$ (see Fig. {\ref{Fig2}}(b)). Then, in the vicinity of these avoided crossing points, the linear approximation can be taken, i.e., $\varepsilon(t)\approx \mp vt$ with $v = A \omega_s^2 \sqrt{1-[\delta/(A\omega_s)]^2}$, so that the AIM can be utilized.

Different from LZ transition of the one-body problem, the cold atoms in the OLC platform are trapped in each lattice site and follow the Boltzmann distribution in the external motional energy levels. Then, the temperature will cause dephasing because the atoms in different motional states follow different LZ transitions. Therefore, the asymptotic population $P_{\rm LZ}$ should be modified by thermally averaging with the Boltzmann-weighted superposition of all external states, that is
\begin{equation}
\langle P_{\rm LZ}\rangle_T = \sum_{n_z,n_x} q(n_z)q(n_x) \exp(-\frac{\pi g_{n_z,n_x}^2}{2v}),               \label{71}
\end{equation}
where $q(n_{z,x})$ are Boltzmann factors in different directions, given in Eq. (\ref{s6}). We also define $\langle \mathcal{O}\rangle_T=\sum_{n_z,n_x}\mathcal{O}(n_z,n_x) q(n_z)q(n_x)$ hereafter with $\mathcal{O}(n_z,n_x)$ is the observable of external state $|n_z,n_x\rangle$ (see Appendix \ref{apdx1}). In order to simulate the experiment, we choose the parameters same as the 1D $^{87}$Sr OLC system \cite{YinCPL2021}, i.e., the longitudinal (transverse) frequency is $\nu_z = 65$ kHz ($\nu_x=250$ Hz), the number of state in longitudinal (transverse) direction is $N_z = 5$ ($N_x=1300$), and the misalignment angle is $\Delta\theta = 10$ mrad. In order to quantitatively verify the temperature effect, we also adopt the numerical Floquet method (see Appendix. \ref{apdx2}) \cite{EckardtRMP2017}.
The population probabilities of state $|e\rangle$ as the function of time $t$ are analyzed in both (a) fast-passage limit at $A=13.3$, $g/\omega_s=0.6$ and (b) slow-passage limit at $A=21.8$, $g/\omega_s=5$. As shown in Fig. \ref{Fig3}, the numerical results (solid lines) are approaching to the asymptotic LZ results $1-\langle P_{\rm LZ}\rangle_T $ (dashed lines). In both cases, a clear jump can be observed around the diabatic states crossing point $t_1=T/4$ with zero detuning. Due to the time interval of the two successive LZ transitions at $t_1$ and $t_2$ is $T/2$, the LZ jump time (the region between the black dotted lines) $\langle t_{\rm LZ} \rangle_T \simeq 0.15T$, which is smaller than $T/2$, is consistent with the step-like structures although it is strictly not a instantaneous moment. Meanwhile, we can find increasing the temperature can strongly depress the excitation, and it is because the Rabi frequency in higher external energy levels is smaller (see Appendix \ref{apdx1}).
On the other hand, the different evolution behaviors are apparent in slow- and fast-passage limits. The high probability of excitation in the slow-passage limit demonstrates the adiabatic process, whereas near zero value in the fast-passage limit reflects the non-adiabaticity. Meanwhile, the first derivatives of $P_e$ in the insets of Fig. \ref{Fig3} present different behavior in different limits, and it may result from state changing in the slow-passage limit. Considering the temperature will not change the step-structure and the transition point, it is set to be 1$\mu$K in the following sections.


\begin{figure}[t]
	\centering
	\includegraphics[width=1\linewidth]{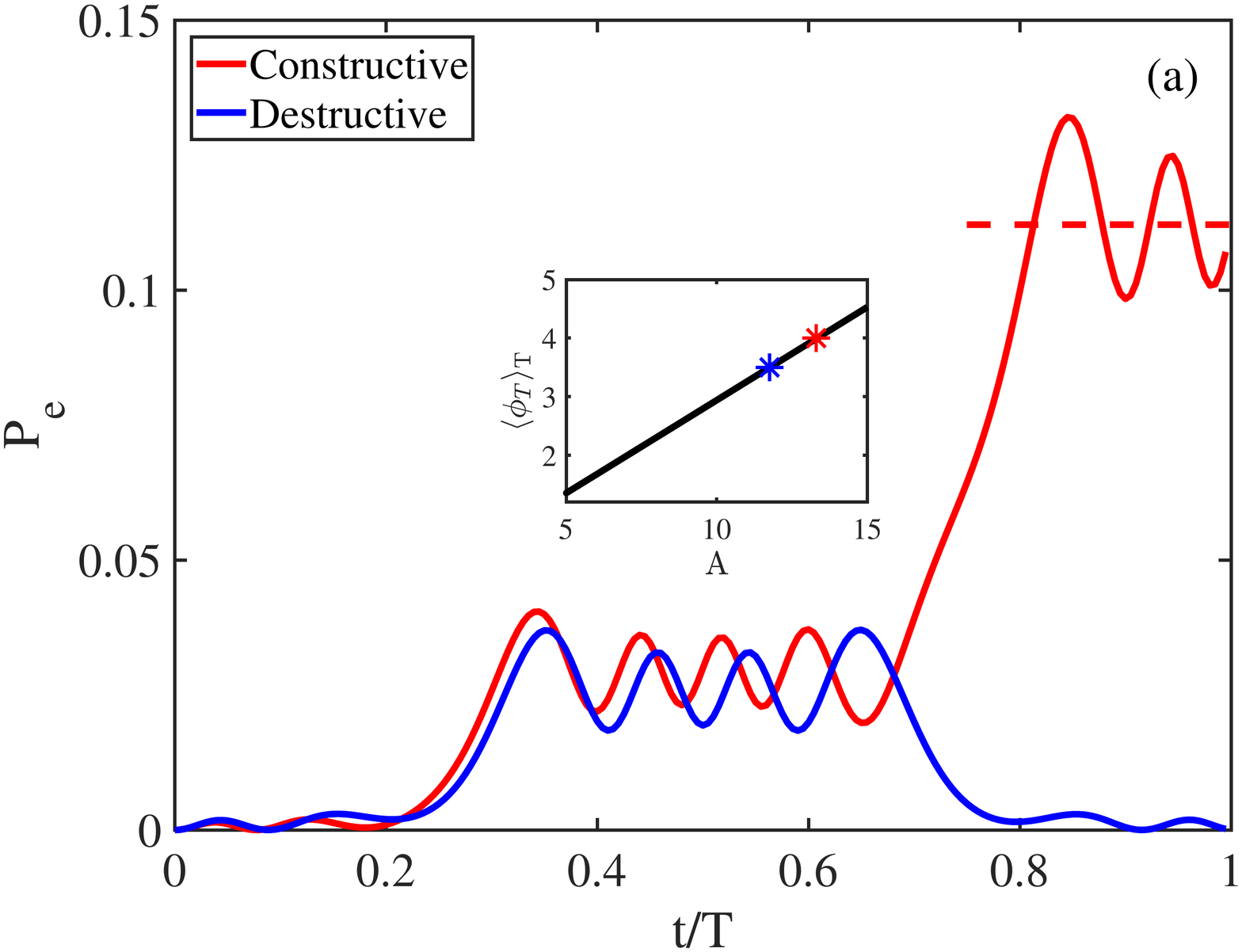}\hfill
	\includegraphics[width=1\linewidth]{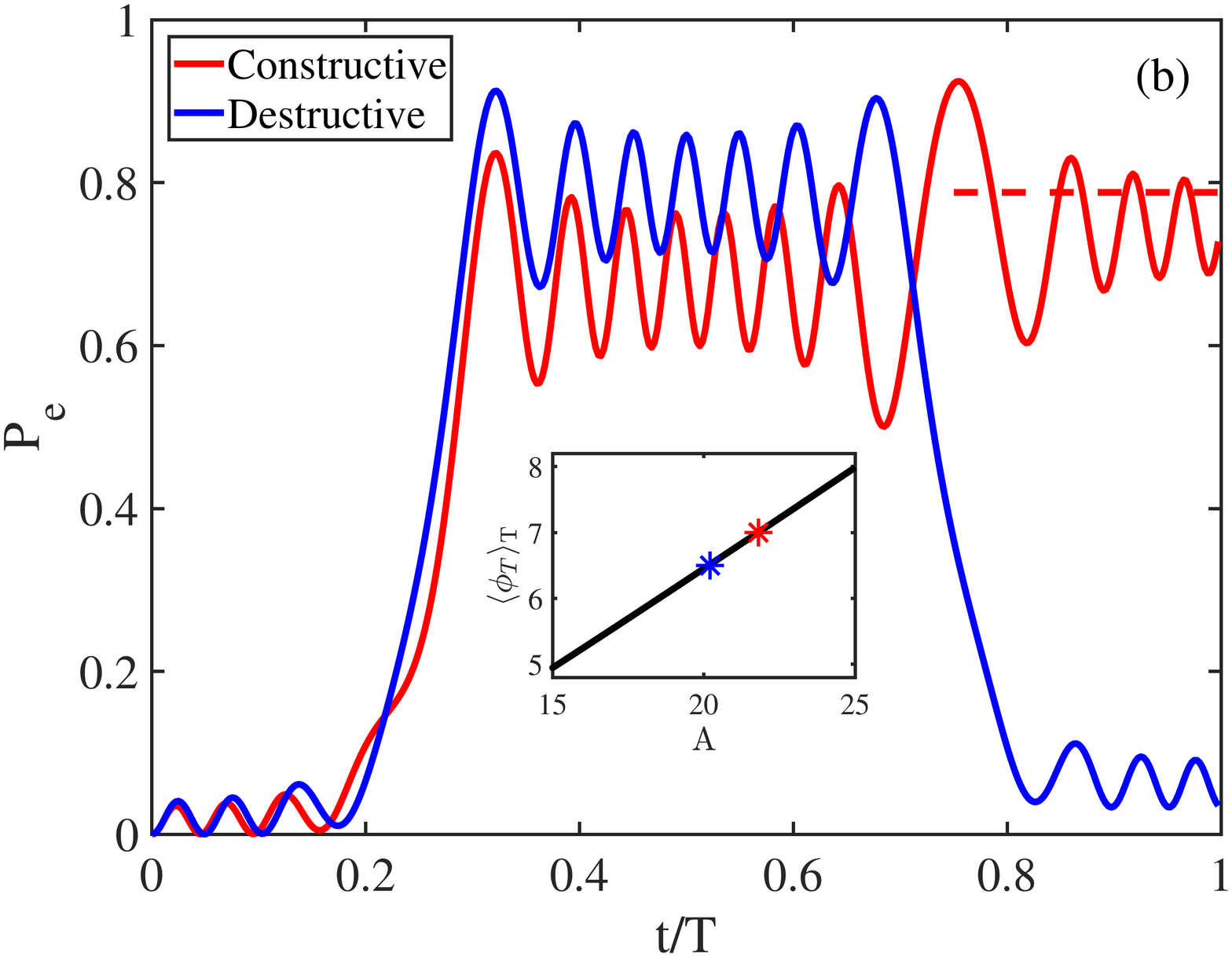}\\
	\caption{Constructive versus destructive interference in one period. The population probability of state $|e\rangle$ is plotted as a function of time with zero detuning in (a) fast-passage limit at $g/\omega_s=0.6$, $A=13.3$ for constructive interference (red lines) and $A=11.75$ for destructive interference (blue lines); (b) slow-passage limit at  $g/\omega_s=5$, $A=21.8$ for constructive interference (red lines) and $A=20.2$ for destructive interference (blue lines). Meanwhile, the dashed lines are $\langle 4P_{\rm LZ} (1-P_{\rm LZ}) \rangle_T$. The insets in both figures are the relationship between the interference phase $\langle \phi_T \rangle_T$ and $A$, where the red (blue) asterisks corresponding to the constructive (destructive) interference. }%
	\label{FigT}%
\end{figure}
\subsection{LZSM Interference in One Period}
Based on the AIM, the LZ transition can be approximately studied via combined evolution matrices. Similarly, as shown in Fig. \ref{0}(b), the evolution matrices at the avoided crossing points $t_1$ and $t_2$ are $X$. Meanwhile, in the other time regions, the corresponding unitary evolution matrix $U_{\phi}$ is with different accumulated phase $\phi$. Then, the whole dynamical process is described by the following matrix
\begin{equation}
U_{T} = U_{\phi_3} X U_{\phi_2} X U_{\phi_1} =
\begin{pmatrix}
\alpha & -\gamma^*\\
\gamma & \alpha^*
\end{pmatrix},                                                                                   \label{a7}%
\end{equation}%
where
\begin{align}
& \alpha  = (1-P_{\rm LZ}) e^{-i\zeta_+} - P_{\rm LZ} e^{-i\zeta_-},                             \label{a8}\\
& \gamma  = -2\sqrt{P_{\rm LZ}(1-P_{\rm LZ})} \cos(\phi_2+\varphi) e^{-i(\phi_1-\phi_3)} ,     \label{a9}
\end{align}
with $\zeta_+  = 2\varphi+(\phi_1+\phi_3)+\phi_2$ and $\zeta_-  = (\phi_1+\phi_3)-\phi_2$, $\phi_i = \phi(t_i,t_{i-1})$, $T = t_3-t_0$.
If assuming the system is prepared in the ground state $|g\rangle$ at $t_0=0$, the population probability of the excited state $|e\rangle$ after one single period can be calculated as
\begin{equation}
P_T = |\gamma|^2 = 4 P_{\rm LZ} (1-P_{\rm LZ}) \cos^2(\phi_2 + \varphi).        \label{a10}%
\end{equation}
We can find $P_T$ is irrelevant to the $\phi_1$ and $\phi_3$ which means only intermediate region between two avoided crossing points takes effect.

The population of excited state oscillates as a function of the interference phase $\phi_T=\phi_2+\varphi$ and it can be interpreted physically as a quantum interference between the two paths along which the system evolves. In comparison with the Mach-Zehnder interferometer in which interference of paths is located in real space, the paths of our system are in phase space. Meanwhile, the avoided crossing acts as a beam splitter and the evolution path acts as the two arms of the Mach-Zehnder interferometer \cite{Science2005,Prlsmerzi2007}. The destructive interference $P_T=0$ happens at $\phi_T=(n+1/2)\pi$, while the constructive one with $P_T = 4P_{\rm LZ} (1-P_{\rm LZ})$ at $\phi_T=n\pi$. For the Hamiltonian Eq. (\ref{0}), $\phi_2$ can be calculated analytically at $\delta=0$ from Eq. (\ref{a5}) \cite{ShevchenkoLTP2006}
\begin{equation}
\phi_2=\frac{\sqrt{A^2\omega_s^2+g_{n_z,n_x}^2}}{\omega_s}  \mathrm{EllipticE}(\frac{A\omega_s}{\sqrt{A^2\omega_s^2+g_{n_z,n_x}^2}}),                \label{phi2}
\end{equation}
where $\mathrm{EllipticE}()$ is the complete elliptic integral of the second kind, so the analytic value of $\phi_T=\phi_2+\varphi$ can be exactly obtained.  In Fig. \ref{FigT}, we show the numerical results of constructive and destructive interference in one period at zero detuning in (a) fast-passage limit with $ \langle \phi_T \rangle_T \simeq 4\pi (7/2\pi)$ for constructive (destructive) interference and (b) slow-passage limit with $ \langle \phi_T \rangle_T \simeq 7\pi (13/2\pi)$ for constructive (destructive) interference. The analytic solution Eq. (\ref{a10}) denotes the maximum of population is $\langle 4P_{\rm LZ} (1-P_{\rm LZ}) \rangle_T$, and the numerical results of constructive interference in both limits can be found oscillating around it (red dashed lines) from Fig. \ref{FigT}. Therefore, with help of Eq. (\ref{phi2}), we can check the relation between the interference phase $\langle \phi_T \rangle_T$ and renormalized driving amplitude $A$. As shown in the inset of Fig. \ref{FigT}, it follows monotonic increase, but also presents nice linearity in large parameter region. It indicates the LZSM interference in OLC platform can be flexibly tuned.

\subsection{Time-domain LZROs for Multi-Period}
One of the most advantages of the ultra-cold $^{87}$Sr OLC platform is its long coherence time, so it is straightforward to expect observing the time-domain LZROs when considering the multiple LZ processes. Following the mathematics of Appendix. B of Ref. \cite{Physrep2010}, the time-evolution over N periods is governed by $N$th power of the single-period evolution operator 
\begin{equation}
 U_T^N = (U_{\phi_3} X U_{\phi_2} X U_{\phi_1})^N.                   \label{a11}
\end{equation}
By diagonalizing the operator $U_{T}$, we can obtain
\begin{equation}
U_T = M E_T M^\dag                                                               \label{a12}%
\end{equation}
with
\begin{align}
M  = \frac{1}{Q}
\begin{pmatrix}
\alpha^*-e^{-i\Phi} & \gamma^*\\
-\gamma             & \alpha-e^{i\Phi}
\end{pmatrix}  ,                                                             \label{a13}
\end{align}
\begin{align}
E_T  =
\begin{pmatrix}
e^{-i\Phi} & 0\\
0          & e^{i\Phi}
\end{pmatrix}  ,                                                              \label{a14}%
\end{align}%
where $Q = \sqrt{|r|^2+|\alpha-e^{-i\Phi}|^2}$, $\cos\Phi = {\rm Re}\alpha$, and thus
\begin{align}
U_T^N & =M E_T^N M^\dag=
\begin{pmatrix}
u_{11} & -u_{21}^*\\
u_{21} & u_{11}^*
\end{pmatrix}         ,                                                       \label{a15} \\
u_{11} & = \cos N\Phi + i({\rm Im}\alpha)\frac{\sin N\Phi}{\sin\Phi} ,           \label{a16}\\
u_{21} & = \gamma \frac{\sin N\Phi}{\sin\Phi} ,                                     \label{a17}
\end{align}
where $\cos\Phi = {\rm Re}\alpha$ \cite{Physrep2010}. Then the population probability of state $|e\rangle$ after N periods is
\begin{equation}
P_e = |u_{21}|^2 = \frac{|\gamma|^2}{|\gamma|^2 + ({\rm Im}\alpha)^2} \sin^2 N\Phi.   \label{a18}%
\end{equation}
We can correspond the integer periods time-dependent factor $\sin^2 N\Phi$ in Eq. (\ref{a18}) to a coarse-grained oscillation $\sin^2\frac{\Omega}{2}t$ with frequency $\Omega=\frac{\omega_s}{\pi}\arccos |{\rm Re}\alpha|$, when $\Omega$ is smaller than the driving frequency $\omega_s$ \cite{PraAshhab2007,GarrawayPRA1997,NeilingerPRB2016}.
If ${\rm Im}\alpha=0$ with solution
\begin{equation}
P_{\rm LZ} \sin\zeta_- = (1-P_{\rm LZ}) \sin\zeta_+                                       \label{a20}
\end{equation}
which is called resonance condition, the amplitude of $P_e$ achieves its maximum value.

In fast-passage limit, $1-P_{\rm LZ} \approx 0^+$, the resonance condition Eq. (\ref{a20}) is proximate $\zeta_-=k\pi$. In the large driving amplitude limit $A\omega_s \gg g_{n_z,n_x}$, we can approximately derive an analytic result \cite{Physrep2010,PraAshhab2007}
\begin{equation}
\zeta_- \simeq \frac{1}{2}\int_0^T \delta+A\omega_s\cos(\omega_st)\,dt = \frac{\delta\pi}{\omega_s},   \label{a22}
\end{equation}
so the resonance condition changes into $\delta = k\omega_s$ which is same as the $k$th Floquet sideband in Ref. \cite{YinCPL2021} with resolved Floquet sideband approximation (RFSA) in the condition $ \omega_s \gg g_{n_z,n_x}$.
Meanwhile, the oscillation frequency can be expressed as
\begin{equation}
\Omega = g_{n_z,n_x}\sqrt{\frac{2}{A\pi}} \cos\left[A-\frac{\pi}{4}(2k+1)\right]                  \label{a24}
\end{equation}
and it will approach to the effective Rabi frequency of $k$th Floquet sideband $g_k^{\rm eff}=g_{n_z,n_x} J_k(A)$ \cite{YinCPL2021} because of the asymptotic behavior of the Bessel functions: $J_k(A) \sim \sqrt{\frac{2}{A\pi}}\cos\left[A-\frac{\pi}{4}(2k+1)\right]$ at $A \gg |k|$.

In the slow-passage limit $P_{\rm LZ} \approx 0$, resonance condition is proximate $\zeta_+=k\pi$. Assuming $A\omega_s \gg g_{n_z,n_x}$ and $\delta=0$, we get $\zeta_+ \simeq 2A - \pi$ \cite{Physrep2010,NeilingerPRB2016}.
Then the resonance condition for constructive interference takes the form $A=\frac{\pi}{2} (2k+1)$ with integer k and the corresponding oscillation frequency is $\Omega = \frac{2\omega_s}{\pi}\sqrt{P_{\rm LZ}}$. Moreover, the resonance condition for destructive interference is $A=k\pi$, which corresponds to $P_T=0$ in Eq. (\ref{a10}).
\begin{figure}[t]
	\centering
	\includegraphics[width=1\linewidth]{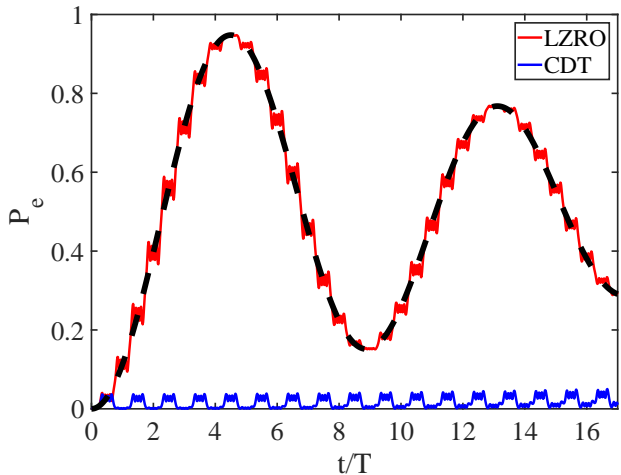}
	\caption{The multiple driving periods in fast-passage limit. The population probability of state $|e\rangle$ is plotted as a function of time with zero detuning. The parameters for red and blue lines are the same as those in Fig. \ref{FigT}(a). The black dashed line shows the coarse-grained oscillation with the Rabi frequency in Eq. (\ref{a24}) for constructive interference. Destructive interference corresponding to CDT is shown with blue solid lines. All the solid lines are numerical results from the Floquet method.}
	\label{Fig4}%
\end{figure}

In order to check the analysis above based on the AIM, we also implement the numerical simulation of long-time evolution. In the fast-passage limit, as shown in Fig. \ref{Fig4}, we plot the population probability of excited state $P_e$ as the function of time t. We can see that $P_e$ exhibits a clear oscillation with step-like structure, and it matches well with the Rabi oscillation with effective frequency given in Eq. (\ref{a24}) (black dashed line). Meanwhile, because of the dephasing effect caused by temperature, the magnitude of $P_e$ is decreasing over time. On the other hand, the blue solid line indicates the destructive interference between diabatic states, which also corresponds to the so-called coherent destruction of tunneling (CDT) \cite{CDTPrl1991}. The CDT phenomena can also be understood with the help of the Rabi frequency Eq. (\ref{a24}). The destructive interference with $A=11.75, k=0$ gives $J_0(11.75) \simeq 0$ and $\Omega\simeq 0$. So the CDT can also be interpreted as the disappearance of the effective Rabi frequency of 0th Floquet sideband \cite{KYPRA1994}.

\begin{figure}[t]
	\centering
	\includegraphics[width=1\linewidth]{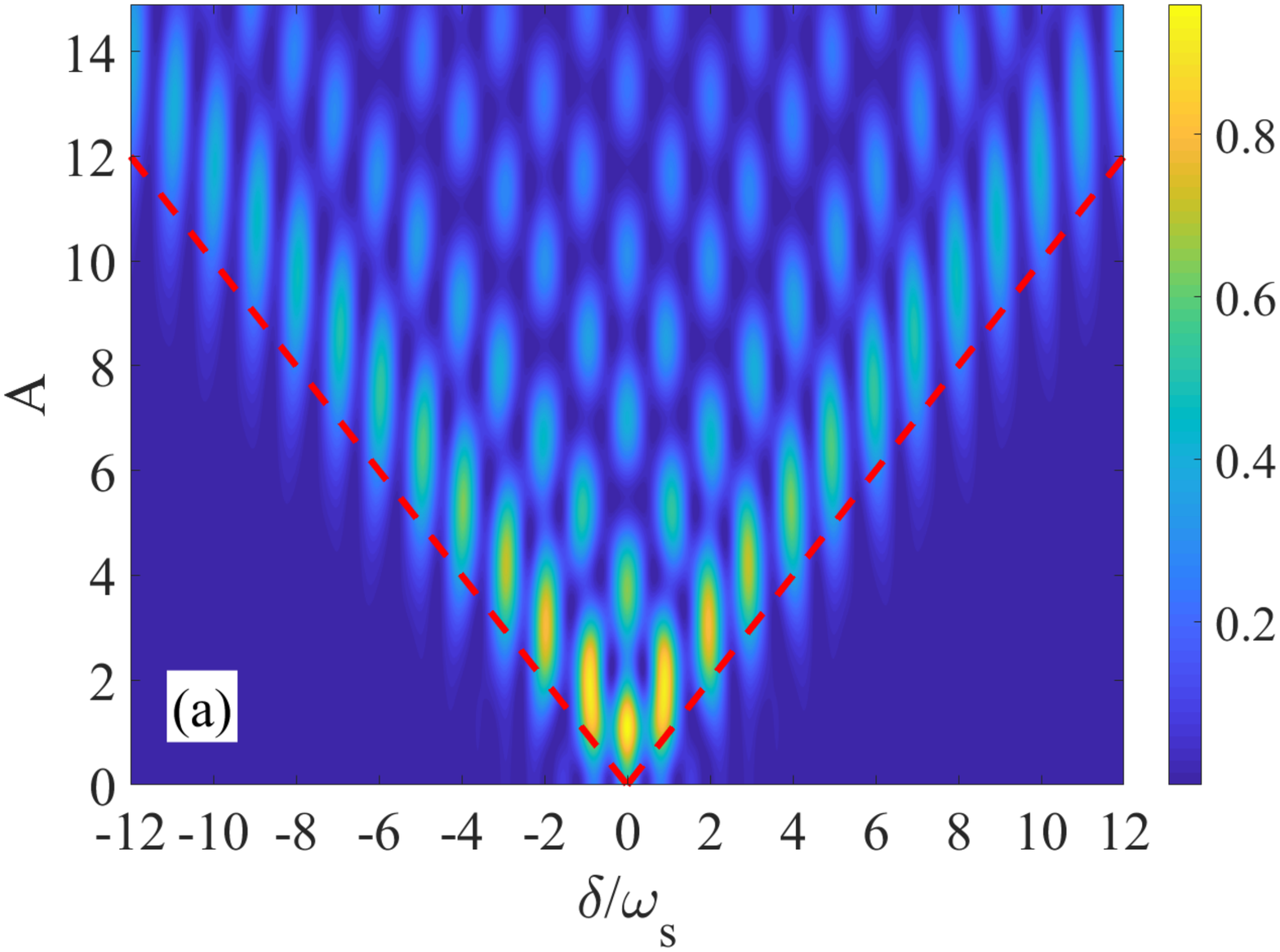}\\
	\includegraphics[width=1\linewidth]{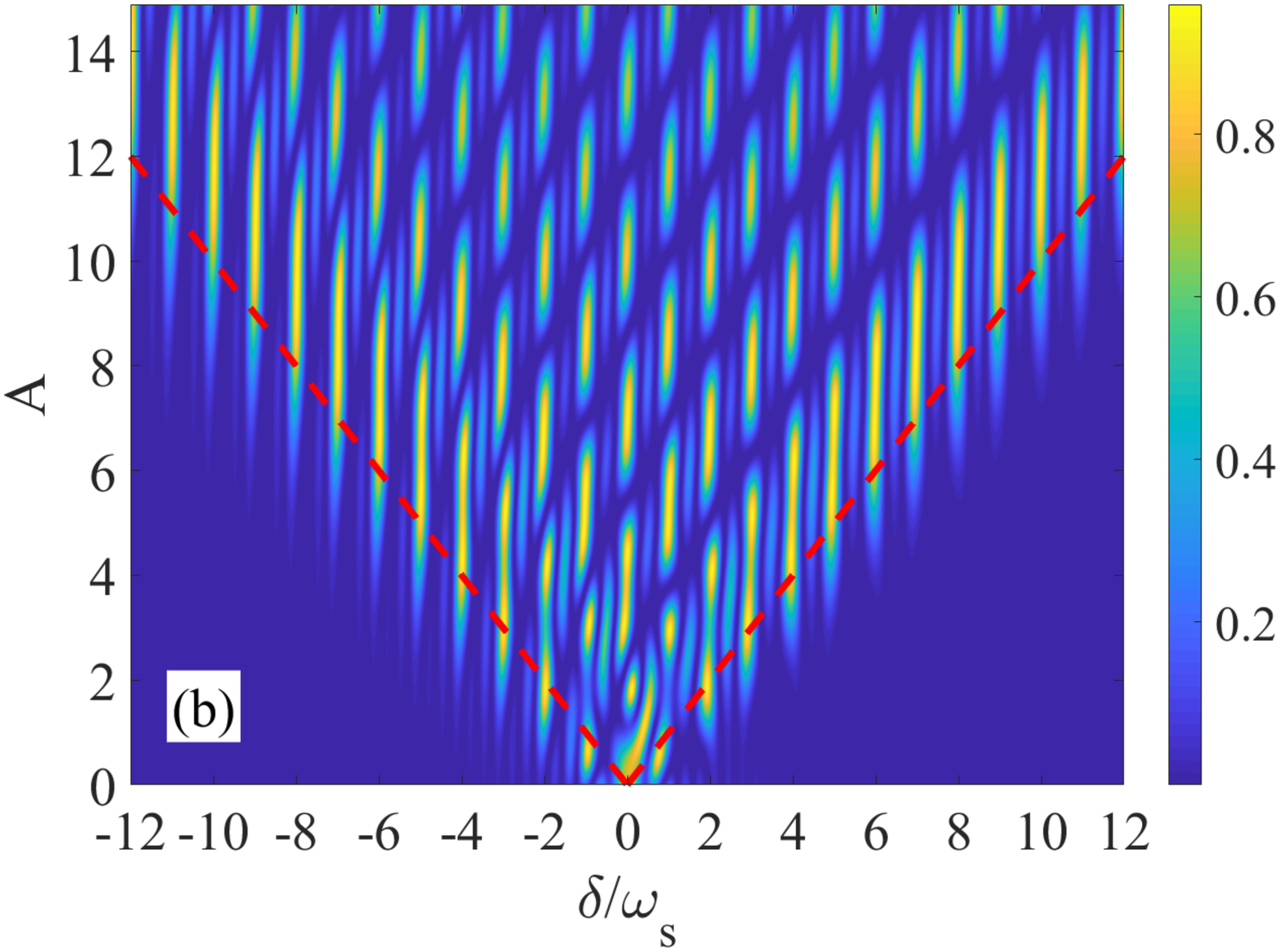}\\
	\caption{The time-domain LZSM interference fringes in the fast-passage limit. The excited-state population probability $P_e$ is plotted as a function of driving amplitude $A$ and the detuning $\delta$ at (a) $t=1.5T$ and (b) $t=3T$  and $g/\omega_s=0.6$. The red dashed lines denote the boundary of the LZSM interference: the region above them indicates the system can experience avoided crossing.}%
	\label{Fig5}%
\end{figure}

Meanwhile, the time-domain interference fringes in the fast-passage limit at a certain driving time are shown in Fig. \ref{Fig5}. Compared with the Fig. 7 of Ref. \cite{Physrep2010}, which is the time-averaged LZSM interference, here we show the interference pattern at different evolution times. We can see that along with the increase of driving amplitude $A$, more interference patterns can be observed, which correspond to more Floquet sidebands being visible. Meanwhile, because of the coarse-grained Rabi oscillation, the probability $P_e$ is monotonic increasing in the first half Rabi period $T_\Omega=2\pi/\Omega$. Thus, stronger interference requires longer evolution time but shorter than $T_\Omega$, as demonstrated in Fig. \ref{Fig5}(b).

\begin{figure}[htp]
	\centering
	\includegraphics[width=0.5\linewidth]{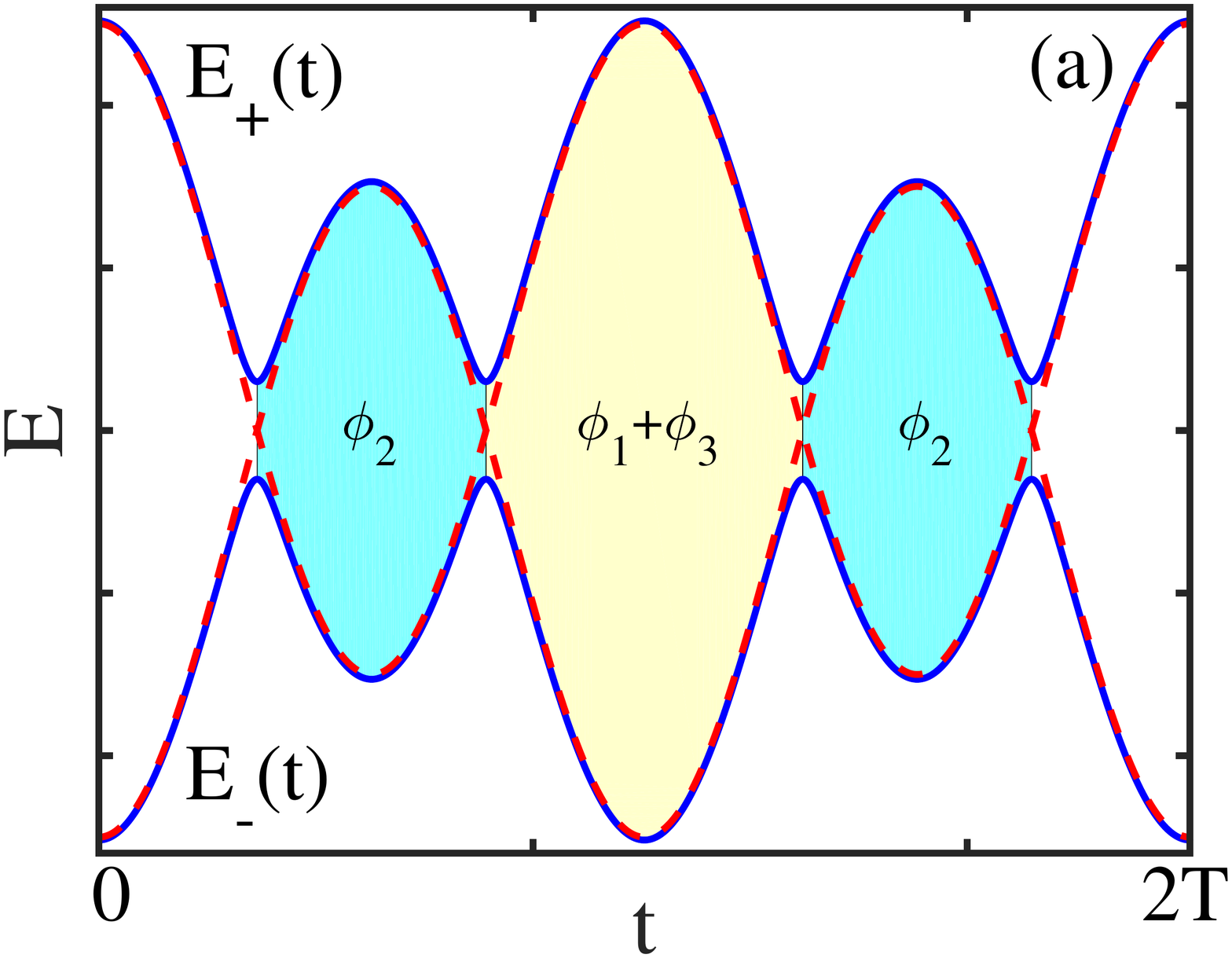}\hfill
	\includegraphics[width=0.5\linewidth]{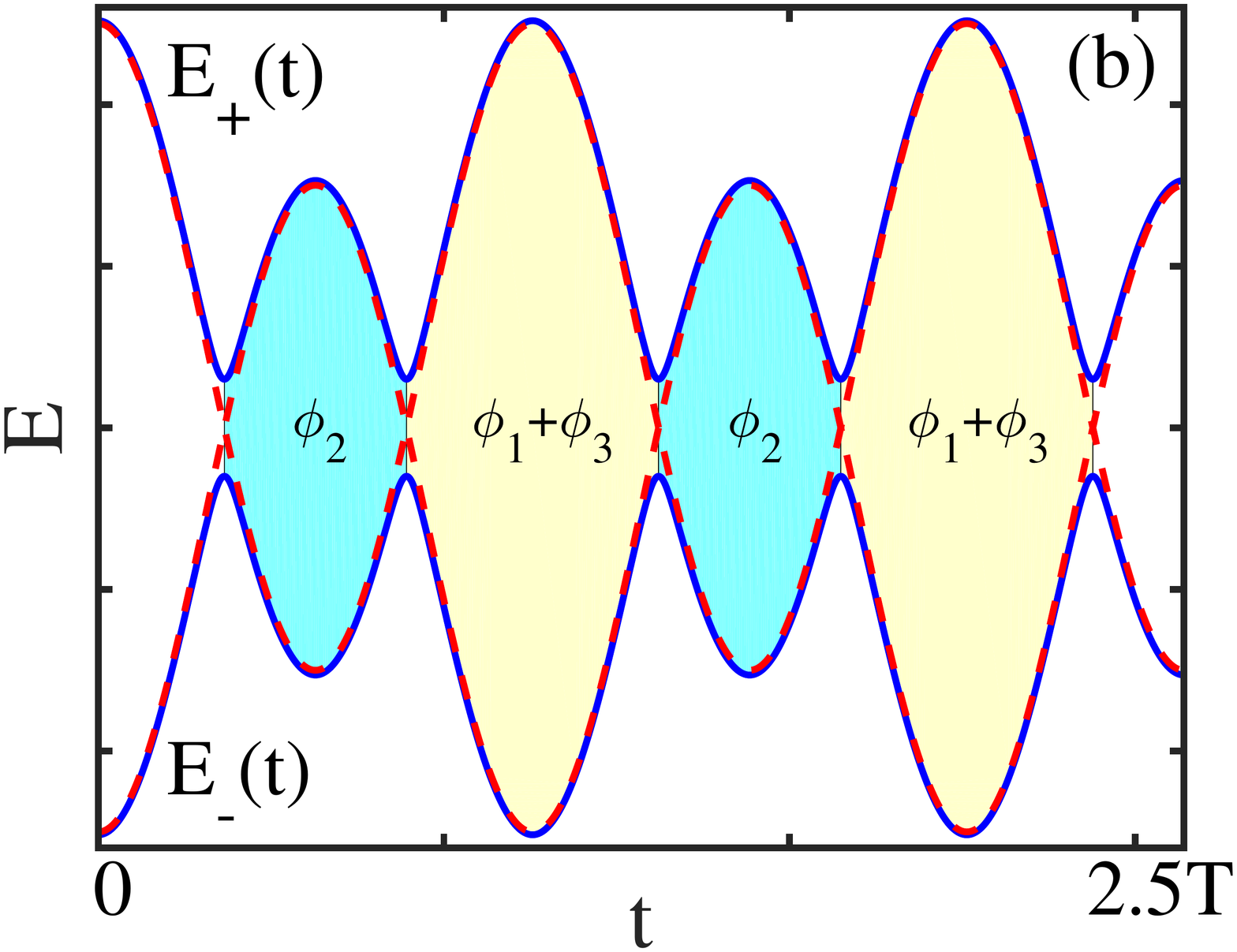}\\
	\caption{Time evolution of the energy levels with (a) $2T$ and (b) $2.5T$. The colored areas indicate the effective dynamic phases $\phi_{1,2,3}$ collected during the adiabatic evolution.}%
	\label{FigIntf}%
\end{figure}

In addition, we find the interference pattern is asymmetry when measuring time is set at integer period ($NT$), but symmetry at half-integer period ($(N+1/2)T$), where $N$ is an integer, as shown in Fig. \ref{Fig5}. This symmetry can be interpreted with the difference of collected dynamic phases during the time evolution. From Fig. {\ref{FigIntf}}, one could see that, when the measuring time is set at $NT$, the picked up effective dynamical phase is $N\phi_2+(\phi_1+\phi_3)(N-1)$ (Fig. \ref{FigIntf}(a)), while at the measuring time $(N+1/2)T$, the picked up effective dynamical phase is $N\phi_2+N(\phi_1+\phi_3)$ (Fig. \ref{FigIntf}(b)). From the Eq. (\ref{a5}) we can see that, when $\delta$ is replaced by $-\delta$, the dynamic phase $\phi_2$ and $\phi_1+\phi_3$ are exchanged with each other. Thus the total effective picked up dynamical phase at measuring time $(N+1/2)T$ will be the same, while at $NT$ it will be different, which accounts for the asymmetry.

\begin{figure}[t]
	\centering
	\includegraphics[width=1\linewidth]{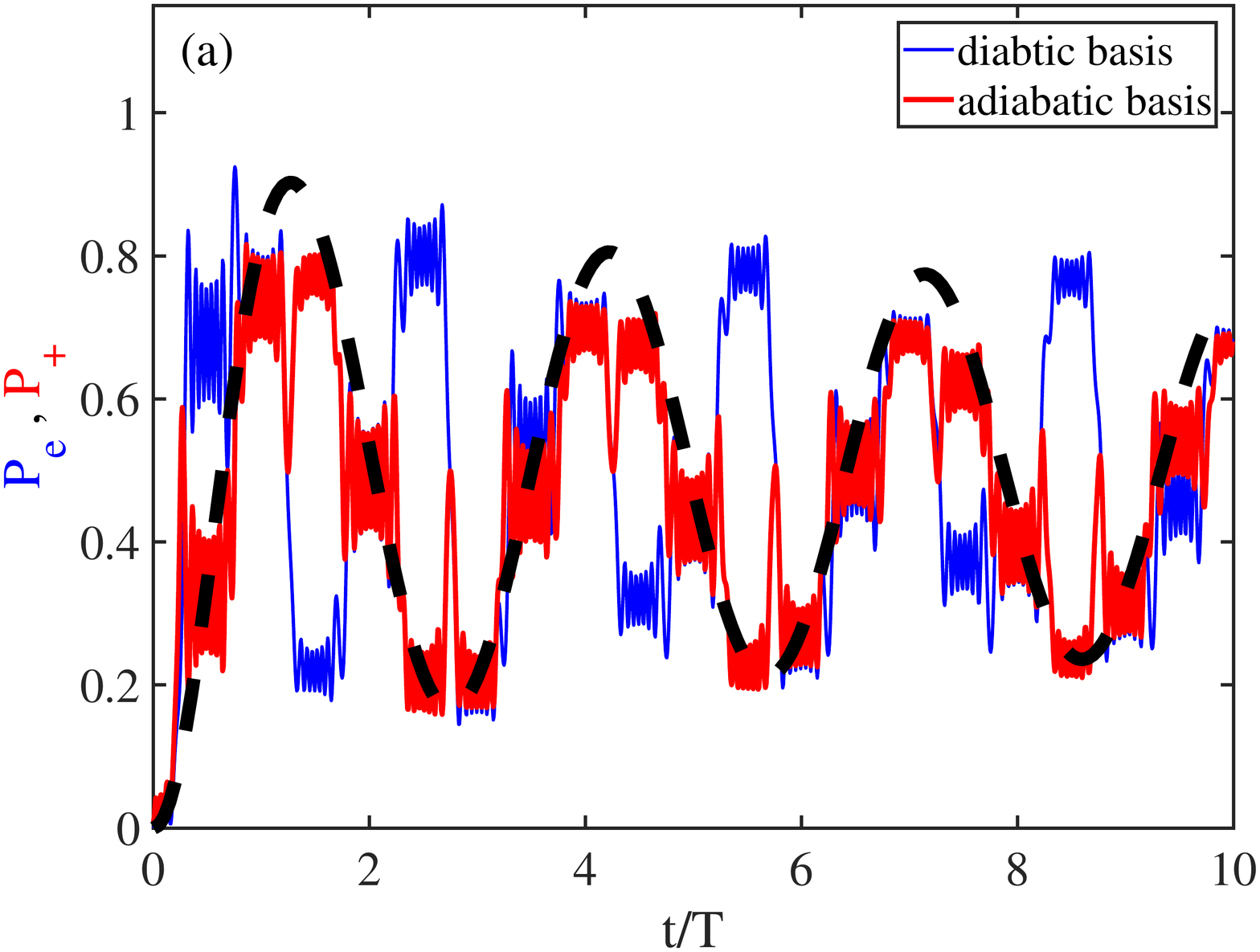}\\
	\includegraphics[width=1\linewidth]{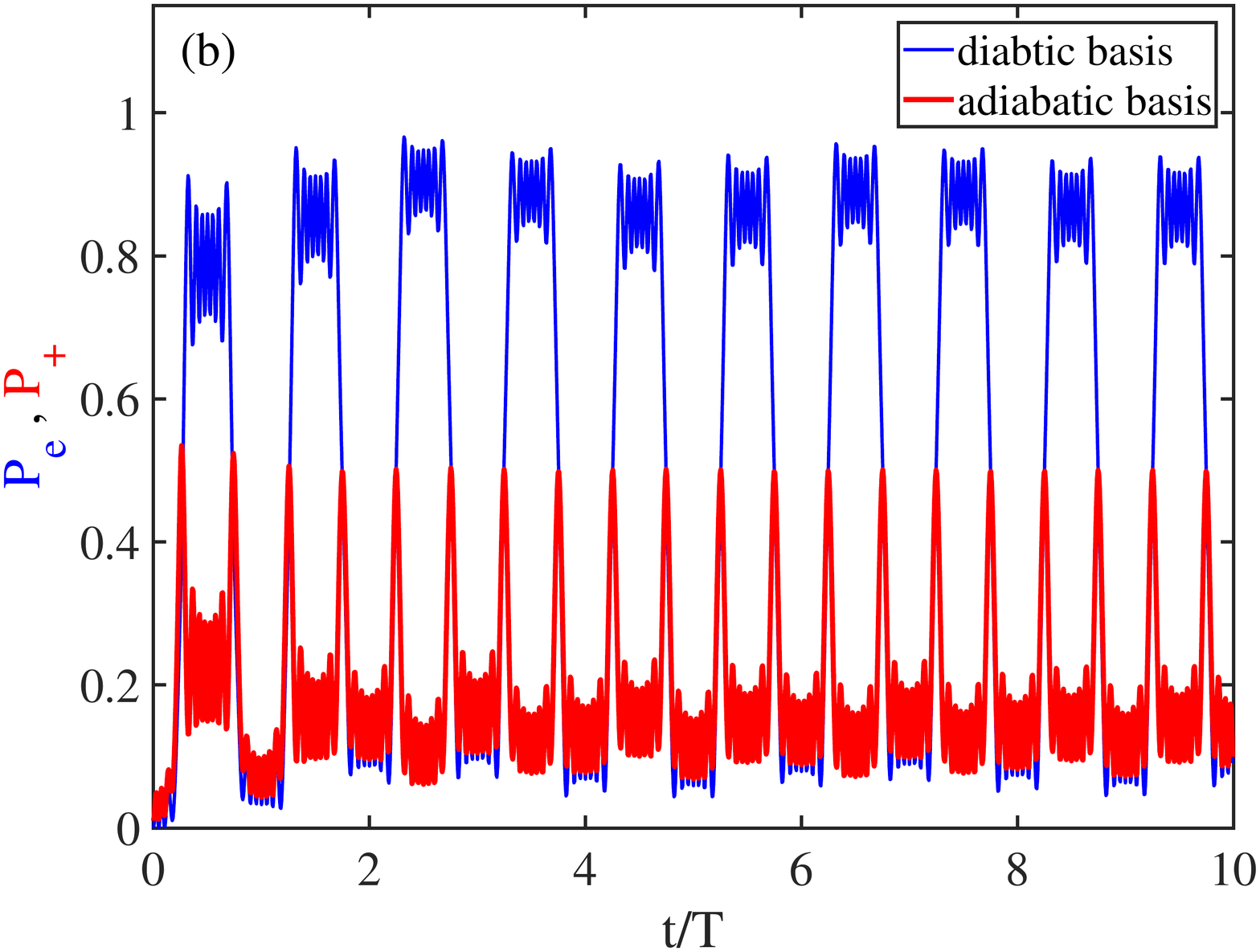}\\
	\caption{The multiple driving periods in slow-passage limit. The population probability of excited state $P_e$ in diabatic basis (blue solid lines) and $P_+$ in adiabatic basis (red solid lines) are plotted as a function of time with zero detuning. The parameters in (a) and (b) correspond to red and blue line in Fig. \ref{FigT}(b), respectively. The black dashed line shows the coarse-grained oscillation with the effective Rabi frequency $\Omega=\frac{\omega_s}{\pi}\arccos|{\rm Re}\alpha|$ for constructive interference. All the solid lines are numerical results from the Floquet method.}
	\label{Fig6}%
\end{figure}

In the slow-passage limit, the population of the excitation $P_e$ follows the coarse-grained oscillation at the integer-period time, but not the half integer ones, as demonstration of Fig. \ref{Fig6}(a). It is because the evolution of the system is almost adiabatic so that the population changes drastically between diabatic states at every avoiding crossing point. Considering the corresponding probability of excited state $P_+$ in the adiabatic basis is small in the first half period, we expect it can follow the AIM result in the whole region. As pointed in the Fig. 5 of Ref. \cite{Physrep2010} in the slow-passage limit, here we also plot the excited probability $P_+$ of the state $|\varphi_+(t)\rangle$ versus the time t.
In Fig. \ref{Fig6}, we compare the population probability in both diabatic and adiabatic basis for both constructive and destructive interference. For the constructive case shown in Fig. \ref{Fig6}(a), we can see the step-like structure associated with multiple LZ transitions, but the long-time dynamics are irregular in diabatic basis. In contrast, after transferring to the adiabatic basis, we see clearly that all the steps follow the LZ Rabi oscillations (black dashed line) including the half period.
In fact, considering the constructive interference under resonance condition Eq. (\ref{a20}) and zero detuning, the population of adiabatic state $P_+$ at N $+1/2$ periods can be expressed as $P_+'=\sin^2 N\Phi+P_{\rm LZ}\cos 2N\Phi+\sqrt{P_{\rm LZ}(1-P_{\rm LZ})}\sin 2N\Phi$ with the combined evolution operator $U_{\phi_2/2}XU_{\phi_1}U_T^N$. Considering the small $P_{\rm LZ}$ in the slow-passage limit, we can approximate $P_+'$ as $\sin^2 N\Phi$, which gives the same Rabi oscillation frequency as the N periods results.

\begin{figure}[htp]
	\centering
	\includegraphics[width=1\linewidth]{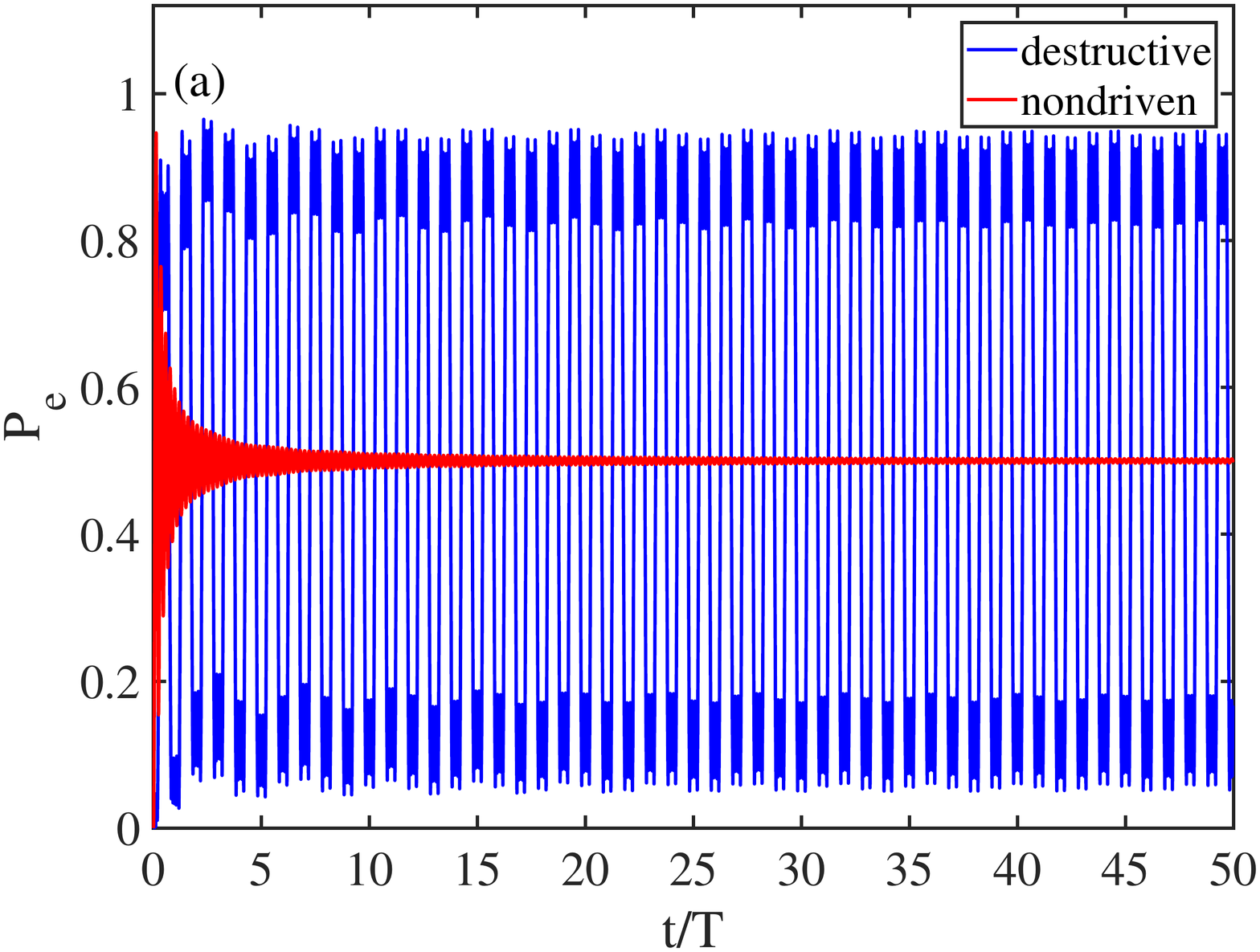}\\
	\includegraphics[width=1\linewidth]{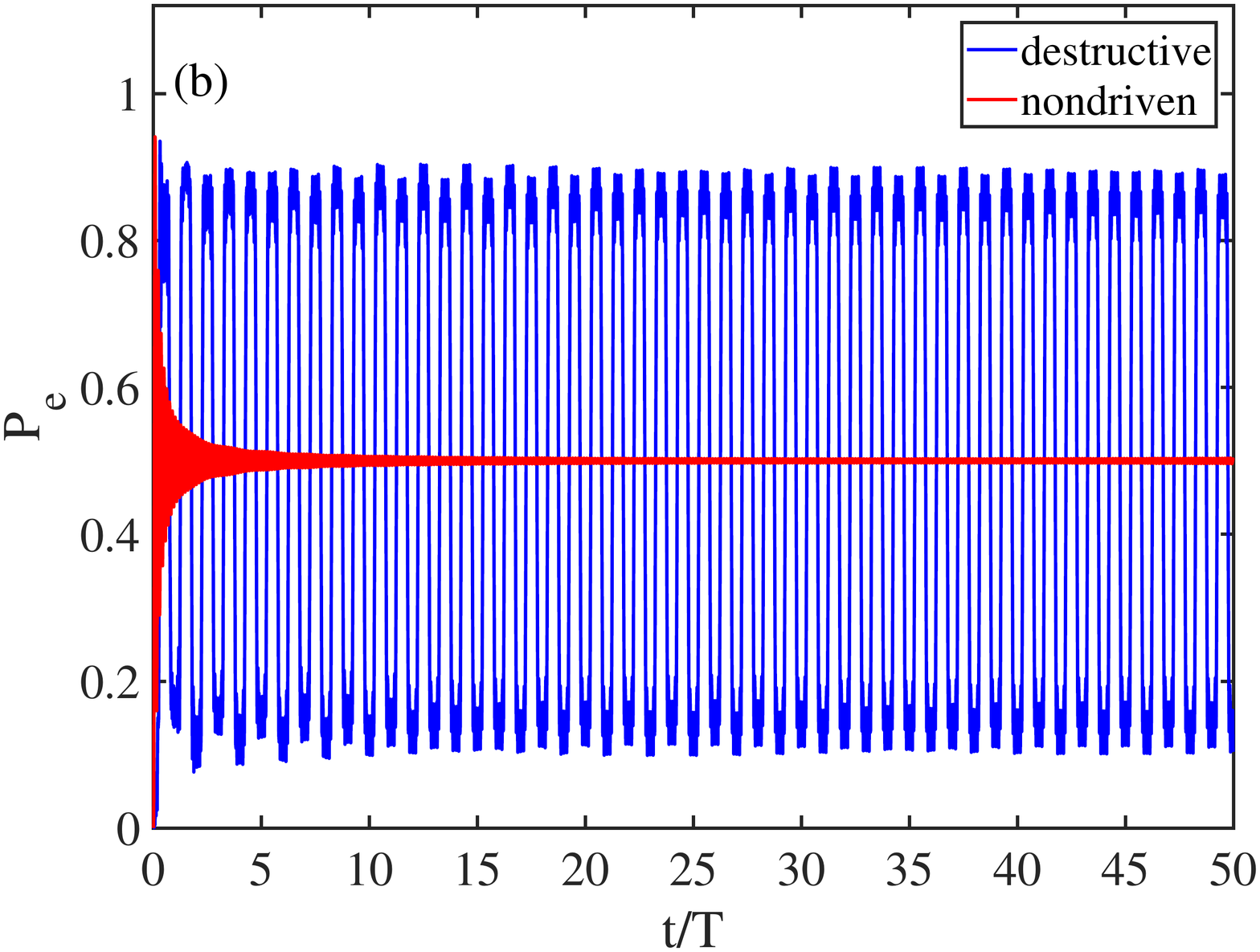}\\
    \includegraphics[width=1\linewidth]{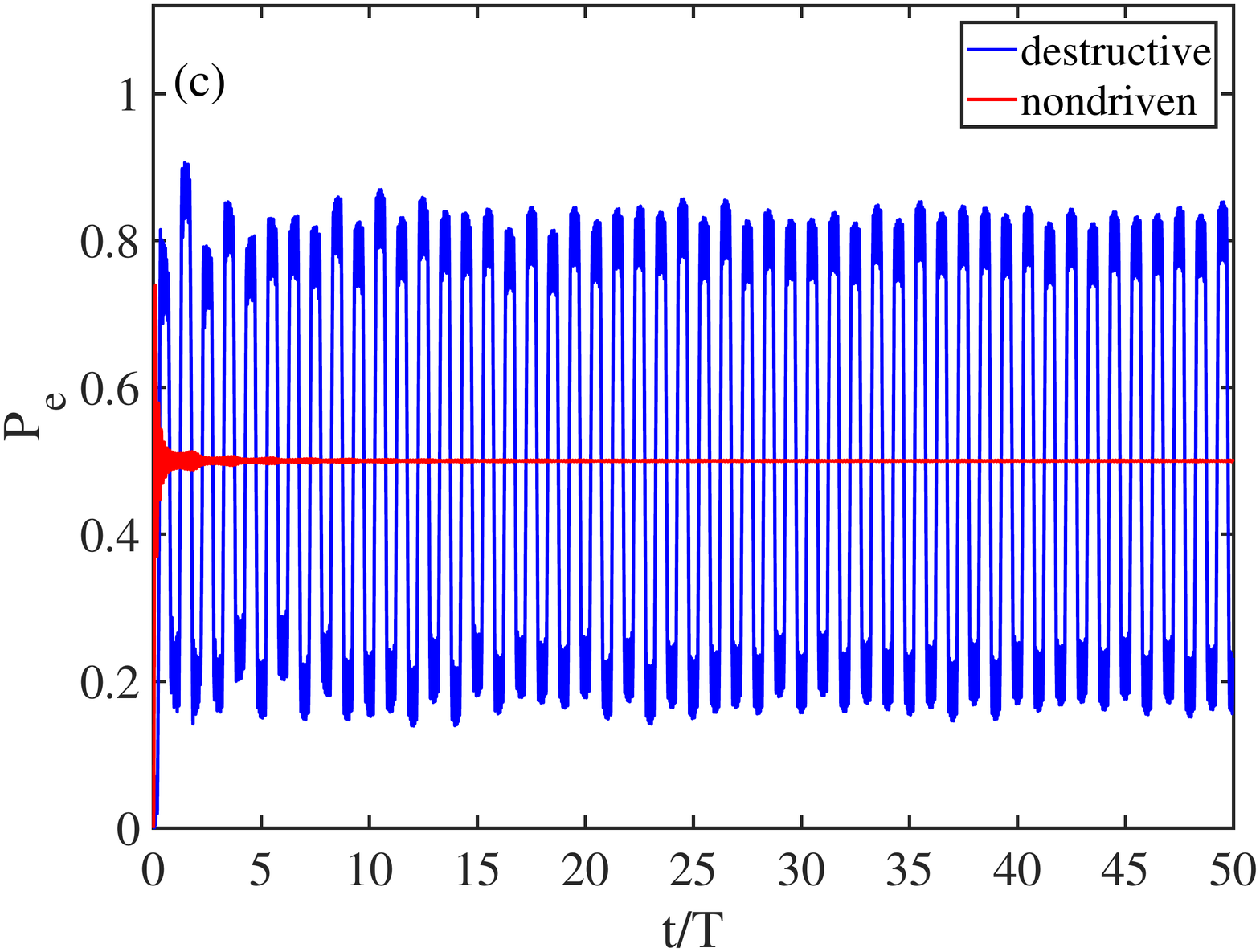}\\
	\caption{Destructive interference in long evolution time for slow-passage limit compared with the non-driven time evolution. The population probability of excited state $|e\rangle$ are plotted with time. The parameters in (a) are the same as those of blue line in Fig. \ref{Fig6}(b) except $A=0$ for red solid line. And the parameters for blue (red) solid line in (b) are chosen with $g/\omega_s=8$, $A=28.62$ ($A=0$), and temperature $T_z=T_x=1\mu$K. The parameters for blue (red) solid line in (c) are chosen with $g/\omega_s=8$, $A=19.56$ ($A=0$), and temperature $T_z=T_x=3\mu$K. All the solid lines are numerical results from the Floquet method.}
	\label{Fig66}%
\end{figure}

For the destructive case shown in Fig. \ref{Fig6}(b), the effect CDT is visible in the adiabatic basis. In comparison, the population probability of excited state in diabatic basis $P_e$ also shows CDT at integer period, but has high amplitude and also the time-domain step-like structures at half-integer period. Meanwhile, interestingly, no apparent dephasing is observed. In order to verify it, we prolong  the evolution time up to fifty-period shown in Fig. \ref{Fig66}(a). We can see the magnitude of $P_e$ (see blue solid line in Fig. \ref{Fig66}(a)) is nearly one without any decay even after fifty driving periods. For comparison, we also plot the population probability of excited state $|e\rangle$ (red solid line in Fig. \ref{Fig66}(a)) under the non-driven situation (that is, $A=0$), which decay to $0.5$ very fast. We also show this non-dephasing effect by enlarging the coupling strength $g$ and elevating the system temperature. In Fig. \ref{Fig66}(b) we plot the population probability of excited state $|e\rangle$ with blue (red) solid line with $g/\omega=8$, $A=28.62$ ($A=0$) at the system temperature $T_z=T_x=1\mu$K, corresponding to the destructive interference (blue solid line) with $\langle \phi_T \rangle_T \simeq 19/2 \pi$; in Fig. \ref{Fig66}(c) we plot the population probability of state $|e\rangle$ with blue (red) solid line with $g/\omega=8$, $A=19.56$ ($A=0$) at the system temperature $T_z=T_x=3\mu$K, corresponding to $\langle \phi_T \rangle_T \simeq 13/2 \pi$ (blue solid line). From both of them we still can not observe any signature of dephasing effect for destructive interference. These results imply that the destructive interference in slow-passage limit may suppress the dephasing effect of clock transition caused by the temperature.

\section{Frequency shift}
\begin{figure}[t]
\centering
\includegraphics[width=1\linewidth]{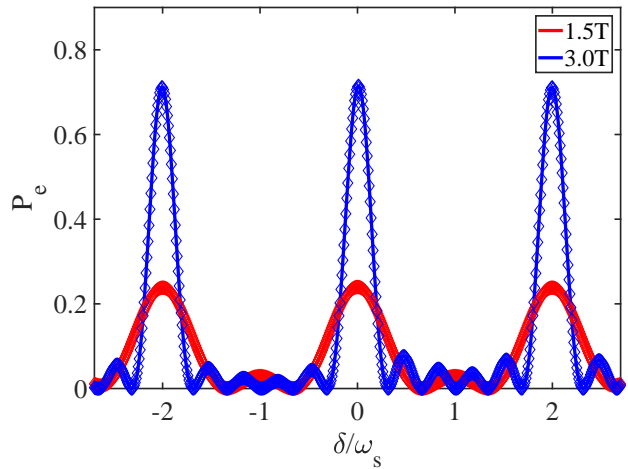}\\
\caption{The population probability of $|e \rangle$ as the function of detuning at $t=1.5T$ (red line) and $t=3T$ (blue line) with $A=13.3$, $g/\omega_s = 0.6$ at 1$\mu$K. The solid lines are the numerical results from Floquet approach, and the colored diamonds indicate the results from AIM.}%
\label{Fig7}%
\end{figure}

Based on the simulation and analysis above, the time-domain LZRO could be well observed in real experiment parameter regions. However, the step-like structure may cause the frequency shift, so that the zero detuning can not be fixed. In order to check it, we simulated the excitation probability as a function of detuning at (a) $t=1.5T$ and (b) $t=3T$ with $A=13.3$, $g/\omega_s=0.6$ and $1\mu$K.
In Fig. \ref{Fig7}, we can see the excitation probabilities for zero detuning are well recognized and thus could be easily chosen to fix the detuning of clock laser.

\begin{figure}[t]
\centering
\includegraphics[width=1\linewidth]{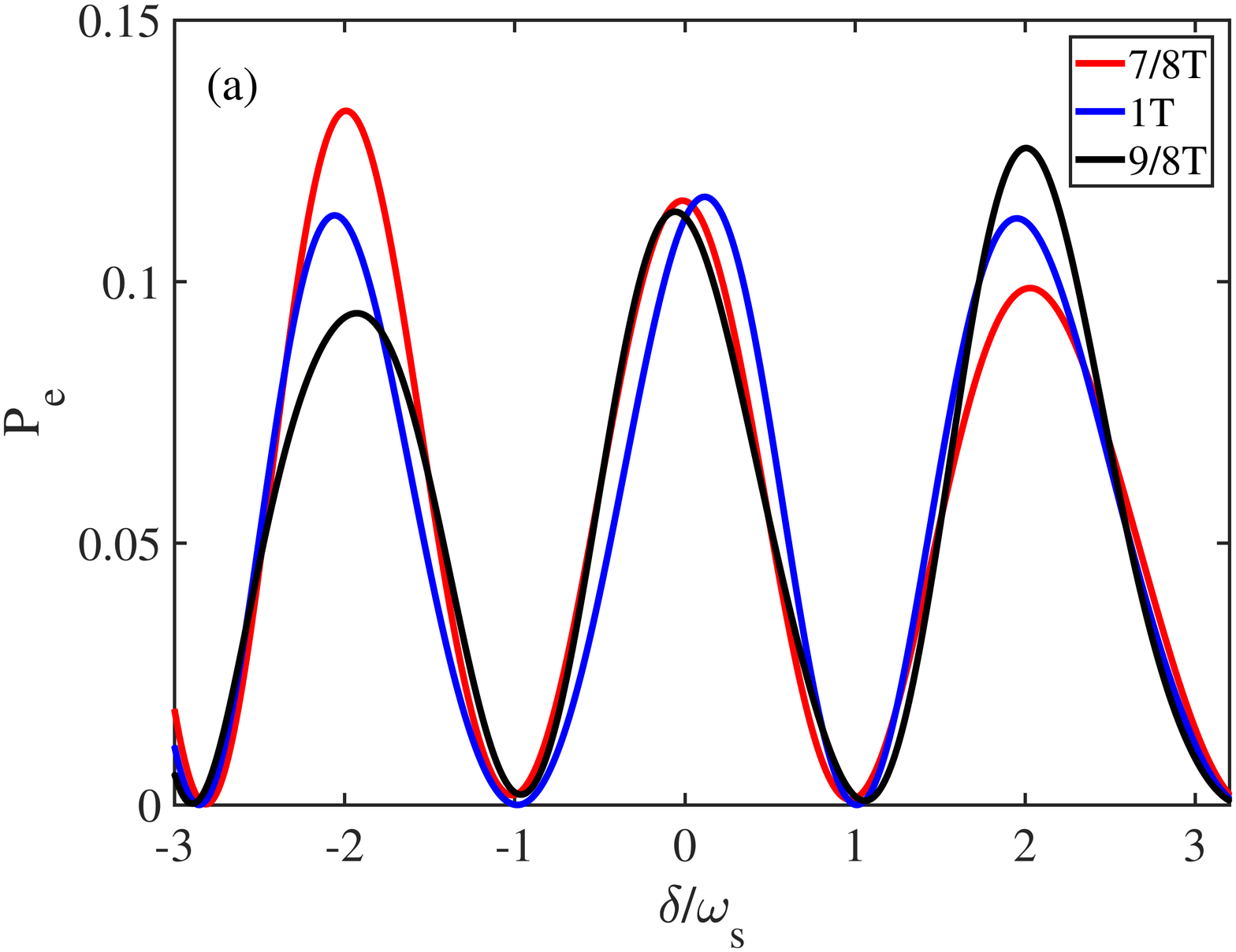}\\
\includegraphics[width=1\linewidth]{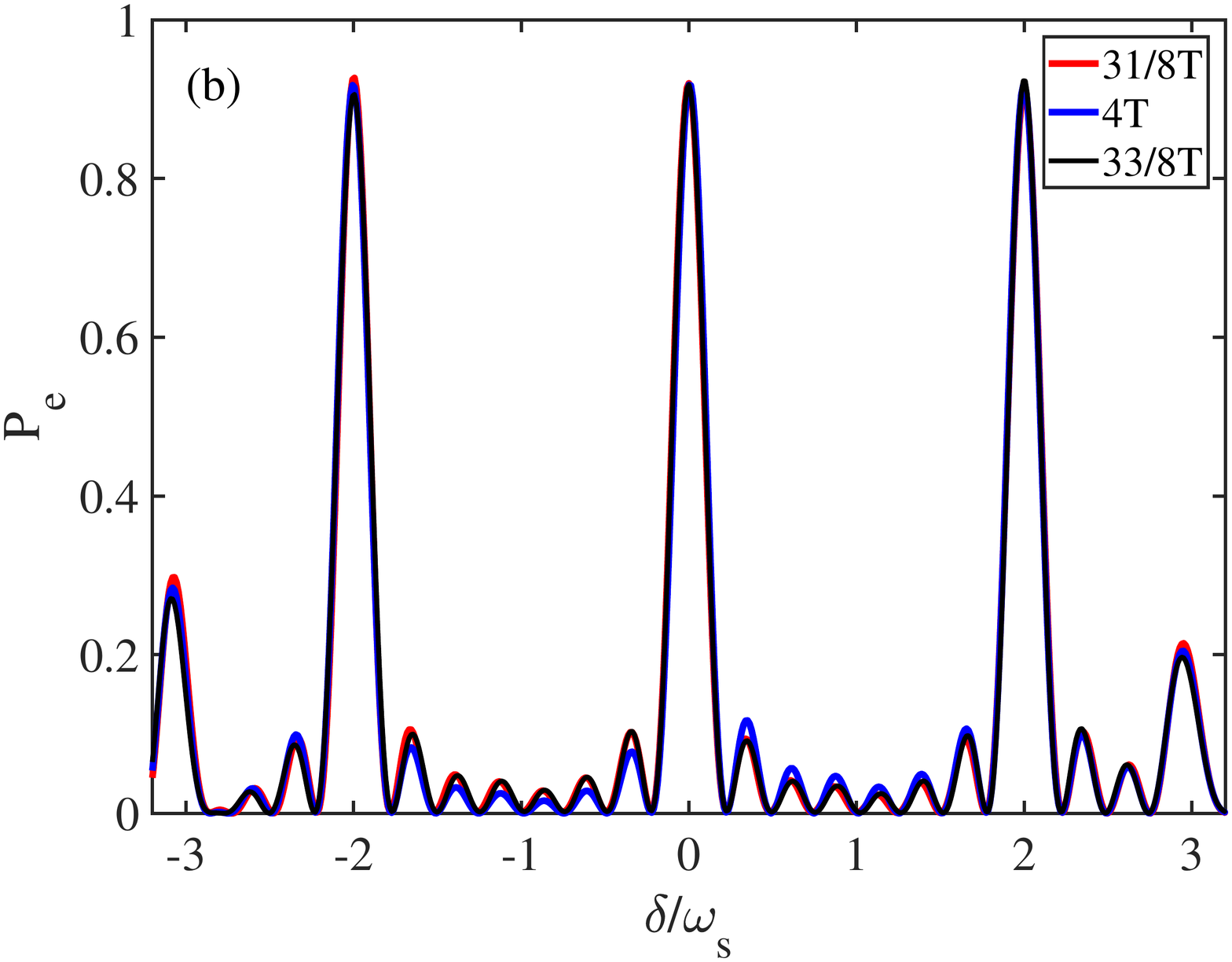}\newline
\caption{The population probability of $|e \rangle$ as the function of detuning near one-period driving time (a) and four-period driving time (b) with $A=13.3$, $g/\omega_s=0.6$ at $1\mu$K, calculated with numerical Floquet approach.}%
\label{Fig8}%
\end{figure}

However, if we consider the pulse duration away from the integer and half-integer period, as shown in Fig. \ref{Fig8}(a) near one-period driving time, there exist obvious shifts in the locations of the peaks corresponding to the generalized Bloch-Siegert shift \cite{TJPrl2010}. In contrast, the spectra near four-period driving time in Fig. \ref{Fig8}(b) do not show any deviations of the peak positions. Thus, the step-like structure of time-domain LZRO should be with distortion at short evolution time, but more obvious at longer evolution time in the fast-passage limit. In addition, here we only consider the region where the driving amplitude is much larger than the coupling strength, that is, $A\omega_s \gg g_{n_z,n_x}$ due to these reasons, (i) small $A\omega_s$ will give large LZ jump time $t_{\rm LZ}$, and then can deteriorate the validity of AIM; (ii) when $A\omega_s$ is large enough, the diabatic and adiabatic states can be treated coincide except in a small region around the avoided crossing. This is not only consist with AIM but also convenient for the further experimental treatment, because one can transform the state populations from the diabatic basis (the measurement basis) to the adiabatic basis more straightforwardly \cite{DuPrl2014}.

\section{CONCLUSION AND DISCUSSION}
With the help of the AIM analytic approach and the numerical Floquet theory method, we study theoretically the periodically modulated one dimensional optical lattice clock system. In both slow- and fast-passage limits, the analytic method matches well with the numerical approach, so that we can locate the proper parameter regions to observe the time-domain LZROs in the real experimental region. In the fast-passage limit with zero detuning, if the constructive condition is fulfilled, the time-domain LZROs follow the coarse-grained oscillation with effective frequency which can be directly obtained from the analytic formula. Meanwhile, the CDT effect can be clearly observed if the destructive interference condition satisfies. On the other hand, in the slow-passage limit, the coarse-grained oscillation curve is proved to be more in accord with the population probability of excited state in the adiabatic basis, which is quite different from the fast-passage limit case. In particular, the long time evolution in slow-passage limit demonstrates the dephasing effect is suppressed with destructive interference.

All the time-domain LZROs we simulated theoretically here are feasible in the real experiment parameter region \cite{YinCPL2021}, so it can be directly used for guiding the experiment. Based on our discussion of the frequency shift, the time-domain LZROs is more easily detectable in the fast-passage limit with long time evolution. Our work not only pave the way for observation of LZROs in the OLC platform, but also shed a light on the quantum metrology based on the LZSM interference in the atomic system. Additionally, the frequency shift discussed in Sec.V opened an interesting issue to be explored about the generalized BS shifts in the OLC system in the future. Moreover, our theory can build a fundamental theoretical framework, serve as the latter for adapting those methods to other situations of interest, such as more complicated optical lattice geometries, or different forms of modulation.

\section{ACKNOWLEDGMENTS}

This work is supported by Special Foundation for theoretical physics Research Program of China (No. 11647165) and China Postdoctoral Science Foundation Funded Project (Project No. 2020M673118). X.-F. Z. acknowledges funding from the National Science Foundation of China under Grants No. 11804034, No. 11874094 and No.12047564, Fundamental Research Funds for the Central Universities Grant No. 2020CDJQY-Z003 and 2021CDJZYJH-003. W.-D. L. acknowledges the funding from the National Natural Science Foundation of China under Grant No. 11874247, the National Key Research and Development Program of China, Grant No. 2017YFA0304500, and the Program of State Key Laboratory of Quantum Optics and Quantum Optics Devices, China, Grant No. KF201703, and the support from Guangdong Provincial Key Laboratory, Grant No. 2019B121203002.

\appendix

\section{Hamiltonian of the driven OLC system}\label{apdx1}

Here we will give a brief introduction to how we treat the driven OLC with both the internal and external degrees of freedom. One can also get a more detailed derivation in the Supplemental material of Ref. \cite{YinCPL2021}.

Under the lattice laser modulation $\omega_L(t)=\bar{\omega}_L+\omega_a\sin(\omega_st)$, the intensity of the lattice laser along the $z$ direction could be described as $I=I_0\sin^2[\bar{\omega}_L(z+\int v(t)dt/c)]$, where $v(t)\simeq\omega_a\omega_sL\cos(\omega_st)/\bar{\omega}_L$ is the effect lattice velocity, $L=0.3$ m denotes the distance between the center of the MOT and the HR mirror. So in the lattice co-moving frame, the frequency of the clock laser that the atom feels needs to shift to $\omega_p'\simeq[1-v(t)/c]\omega_p$ duo to the relativistic Doppler effect. This modulating pattern does not change the external trapping potential. Considering a small misalignment angle $\Delta\theta$ along the transverse $x$-direction (we can choose the $\Delta\theta$ along $x$-axis because the transverse trap is isotropic) between the clock laser axis and the lattice axis in real experiments, the external Hamiltonian with lattice potential having a Gaussian profile in the $x$ direction can be expressed as%
\begin{equation}
	\hat{H}_{\rm ext}=\frac{\hat{p}^2}{2m}-U_0\sin^2\left(\frac{\bar{\omega}_Lz}{c}\right) e^{-\frac{2x^2}{w_0^2}}, \label{s1}%
\end{equation}
where $U_0=4\alpha_0P_0/(\pi c\varepsilon_0w_0^2)$ is the lattice depth, $x$ denotes the transverse distance from the lattice axis. Due to the large lattice depth $U_0$, we can neglect the intersite tunneling. With the harmonic approximation, the external Hamiltonian can be expressed as%
\begin{equation}
	\hat{H}_{\rm ext}\simeq\frac{\hat{p}^2}{2m}-\left(\frac{U_0 \bar{\omega}_L^2}{c^2}z^2+\frac{2U_0}{w_0^2}x^2\right). \label{s2}%
\end{equation}
This external Hamiltonian has the harmonic oscillator eigenstates $|n_z,n_x\rangle$ with the corresponding eigenenergies%
\begin{equation}
	E_{n_z,n_x}=h\nu_z(n_{z}+1/2)+h\nu_x(n_{x}+1), \label{s3}%
\end{equation}
where $\nu_z=\sqrt{U_0\bar{\omega}_L^2/(2m\pi^2w_0^2)}$, $\nu_x=\sqrt{U_0/(m\pi^2w_0^2)}$ are the longitudinal and transverse trap frequencies, respectively. The number of the harmonic oscillator states in the trap is approximately given by $N_zN_x^2$, where $N_{z,x}\simeq U_0/(h\nu_{z,x})$ is the number of states in the longitudinal (transverse) direction. Meanwhile, the internal Hamiltonian (after the rotating wave approximation) can be written as the LZSM Hamiltonian Eq. (\ref{0}) \cite{YinCPL2021},
in which $\delta = \omega_0 - \omega_p$ is the frequency detuning of the clock laser, $A=\hbar \omega_a \omega_p L/(c \bar{\omega}_L)$ is the renormalized driving amplitude, which is dimensionless and independent of driving frequency $\omega_s$. $g_{n_z,n_x}=g e^{-\eta_z^2/2} e^{-\eta_x^2/2} L_{n_z}(\eta_z^2) L_{n_x}(\eta_x^2)$ is the coupling strength in the external state $| n_z,n_x \rangle$ \cite{BlattPRA2009}, where $L_n$ is the Laguerre polynomial. $\eta_z = \sqrt{h/(2m \nu_z)}/ \lambda_p$, $\eta_x = \sqrt{h/(2m\nu_x)} \Delta \theta/\lambda_p$ are the Lamb-Dicke parameters.

Due to the temperature being a few $\mu$K in the system, we can consider a normalized Boltzmann distribution to describe the cold atoms in the external states $|n_z,n_x\rangle$. Then the clock transition probability can be expressed by%
\begin{equation}
	\langle P_e \rangle_T= \sum_{n_z,n_x} q(n_z) q(n_x) P_e \label{s5}%
\end{equation}
as a Boltzmann-weighted superposition of single external state transition probability $P_e$ obtained from the Hamiltonian Eq. (\ref{0}), where $q(n_z)$ ($q(n_x)$) are the Boltzmann-weights corresponding to the longitudinal (transverse) temperature $T_z$ ($T_x$)%
\begin{equation}
	q(n_{z,x}) = \frac{e^{-E_{n_z,x}/(k_BT_{z,x})}}{\sum_{n_{z,x}}e^{-E_{n_z,x}/(k_BT_{z,x})}} \label{s6}%
\end{equation}%
with $E_{n_{z,x}}=(n_{z,x}+1/2)h\nu_{z,x}$ is the energy of motional state $|n_{z,x}\rangle$, $k_B$ is the Boltzmann constant.

\section{Floquet approach}\label{apdx2}

The dynamics of internal clock transition governed by the time-periodic LZSM Hamiltonian Eq. (\ref{0}) can also be solved by the Floquet theory, which provides an accurate approach to deal with the periodically driven quantum system \cite{Shirly1965,EckardtRMP2017,BMAIP2015}.

According to the Floquet theorem, the quantum system described by the time-periodic Hamiltonian $\hat{H}(t)=\hat{H}(t+T)$, gives rise to generalized stationary states called Floquet states. These Floquet states are the solutions of the periodically time-dependent Schr\"{o}dinger equation%
\begin{equation}
i\hbar\frac{\partial}{\partial t}|\psi(t)\rangle = \hat{H}(t)|\psi(t)\rangle                                  \label{19}%
\end{equation}
and have the form%
\begin{equation}
|\psi_\alpha(t)\rangle = |u_\alpha(t) \rangle e^{-\frac{i}{\hbar}\varepsilon_\alpha t},                        \label{20}%
\end{equation}
where $\varepsilon_\alpha$ are the quasienergies, which are the eigenvalues of the so-called Floquet Hamiltonian $\hat{H}_F=\hat{H}(t)-i\hbar\partial/\partial t$:
\begin{equation}
\hat{H}_F|u_\alpha(t)\rangle = \varepsilon_\alpha|u_\alpha(t)\rangle   \label{21}%
\end{equation}
and $|u_\alpha(t)\rangle$ are the Floquet modes, which have the same periodicity with the Hamiltonian, i.e. $|u_\alpha(t+T)\rangle =|u_\alpha(t)\rangle$.

The Floquet states are also the eigenstates of the time-evolution operator over integer driving periods with the eigenvalues $\exp(-i\varepsilon_\alpha nT/\hbar)$ ($n$ is an arbitrary integer):
\begin{equation}
\hat{U}(t_0+nT,t_0)|\psi_\alpha(t_0)\rangle = e^{-\frac{i}{\hbar}\varepsilon_\alpha nT}|\psi_\alpha(t_0)\rangle, \label{22}%
\end{equation}
$\hat{U}(t_2,t_1)$ is the time evolution operator from $t_1$ to $t_2$. The Floquet states can be computed with the time-evolution operator, $|\psi_\alpha(t)\rangle =\hat{U}(t,t_0)|\psi_\alpha(t_0)\rangle $. Note that the eigenvalue $\exp(-i\varepsilon_\alpha T/\hbar)$ does not depend on the initial time $t_0$. So the time-evolution operator can be composed by these complete and orthogonal Floquet states at any fixed time $t$%
\begin{equation}
\hat{U}(t,t_0) = \sum_\alpha e^{-\frac{i}{\hbar}\varepsilon_\alpha(t-t_0)}| u_\alpha(t)\rangle\langle u_\alpha(t_{0})|. \label{23}%
\end{equation}

Now consider the state
\begin{equation}
|u_{\alpha,n}(t)\rangle = |u_\alpha(t)\rangle e^{in\omega_st}  ,                                                \label{24}%
\end{equation}
which gives physically equivalent state to $|u_\alpha(t)\rangle$, it is also a solution to Eq.(\ref{21}) with the shifted quasienergy $\varepsilon_{\alpha,n} =\varepsilon_\alpha + n\hbar\omega_s$, but the corresponding Floquet state is not altered%
\begin{equation}
|\psi_\alpha(t)\rangle = |u_{\alpha,n}(t)\rangle e^{-\frac{i}{\hbar}\varepsilon_{\alpha,n}t}
= |u_\alpha(t)\rangle e^{-\frac{i}{\hbar}\varepsilon_\alpha t} .                                                          \label{25}%
\end{equation}

Now considering the system is initially in the ground state $|g\rangle$, which means $|\psi_\alpha(t_0)\rangle =|g\rangle$, then the population probability of the state $|e\rangle$ at time $t$ can be obtained by%
\begin{equation}
P_e = |\langle e|\hat{U}(t,t_0)|g\rangle|^2 .                                          \label{26}%
\end{equation}

Then the problem turns to finding out the Floquet modes $|u_\alpha(t)\rangle$, and the corresponding quasienergy $\varepsilon_\alpha$. The Eq. (\ref{23}) constitutes an eigenvalue problem in an extended Hilbert space \cite{Shirly1965,Sambe1973}, which is given by the product space of the original Hilbert space and the time-dependent Fourier space. Here this extended Hilbert space can be constructed by:
\begin{equation}
[|g\rangle,|e\rangle] \otimes [1,e^{\pm i\omega_st}, e^{\pm2i\omega_st}\ldots]                                   \label{27}%
\end{equation}
In this extended Hilbert space,
\begin{equation}
|u(t)\rangle = \sum_{n=-\infty}^{+\infty}|u^n\rangle e^{in\omega_st} ,                           \label{28}%
\end{equation}
where each $|u^n\rangle = (u_p^n,u_s^n)^{\rm T}$ (the ${\rm T}$ here denotes a transpose) is a two-component vector, and the component of the Floquet Hamiltonian $\hat{H}_F$ can be represented as%
\begin{align}
\hat{H}_F^{m-n} & =\frac{\omega_s}{2\pi}\int_0^{\frac{2\pi}{\omega_s}}e^{-im\omega_st}\left[\hat{H}_{\rm LZSM}^{n_z,n_x}(t)-i\hbar\frac{\partial}{\partial t}\right]  e^{in\omega_st}\,dt\nonumber\\
& =\hat{H}_{m-n}+n\hbar\omega_s\delta_{m,n} ,                                                                                       \label{29}%
\end{align}
where $\hat{H}_{m-n}+n\hbar\omega_s\delta_{m,n}\hat{I}$ ($\hat{I}$ is a 2$\times2$ identity matrix) forms the $m$th row and $n$th column of the Floquet block. For clarity, the Floquet Hamiltonian can be visualized in this extended Hilbert space as the following matrix form%
\begin{equation}
\left(
\begin{array}{ccccc}
\ddots & \vdots                  & \vdots                  & \vdots & \\
\cdots & \hat{H}_0-\hbar\omega_s & \hat{H}_{-1}            & 0                       & \cdots\\
\cdots & \hat{H}_{+1}            & \hat{H}_0               & \hat{H}_{-1}            & \cdots\\
\cdots & 0                       & \hat{H}_{+1}            & \hat{H}_0+\hbar\omega_s & \cdots\\
& \vdots                  & \vdots                  & \vdots                  & \ddots
\end{array}
\right),
\end{equation}                                                                                                                      \label{30}%
where%
\begin{equation}
\hat{H}_0 = \frac{\hbar}{2}
\begin{pmatrix}
\delta      & g_{n_z,n_x}\\
g_{n_z,n_x} & -\delta
\end{pmatrix}  ,      \nonumber
\end{equation}%
\begin{equation}
\hat{H}_{+1} = \hat{H}_{-1} = \frac{\hbar}{4}
\begin{pmatrix}
A\omega_s & 0\\
0         & -A\omega_s
\end{pmatrix} ,        \nonumber
\end{equation}
and the blocks with $|m-n|$ larger than 1 are zero. 

Now, we can see that the time-dependent Schr\"{o}dinger equation is transformed into an eigenvalue equation of an infinite dimensional but time-independent Hamiltonian with infinitely repeating block structure. Obviously, this eigenvalue problem has an infinite number of solutions, but we can see that those infinite solutions can be divided into two groups, each of which are related to each other, and the same group of the solutions generate the identical Floquet state by Eq. (\ref{25}). So we can truncate the number of the Floquet blocks that converged sufficiently to get the eigenvalues of this infinite dimensional Hamiltonian, of which just two solutions are needed. In order to get the numerical solutions in Sec.III and IV, $161$ Floquet blocks ($n=-80$ to $80$) are truncated, and we choose the two solutions with $n=0$ to get the transition probability with Eq. (\ref{23}) and Eq. (\ref{26}). It is worthwhile to note that we have not made any approximation to solve the original time-periodic Hamiltonian. This should not be confused with the Floquet-Magnus expansion method which is only applicable in high frequency regime \cite{EckardtRMP2017}. In contrast, the Floquet approach used in our manuscript can be applied to any parameter regime.
\bibliographystyle{apsrev4-1}
\bibliography{referen}

\end{document}